\documentclass[aps,showpacs,showkeys,preprintnumbers,nofootinbib]{revtex4}

\usepackage{amsmath}
\usepackage{graphicx}
\usepackage{bbm}                                                  

\newcommand{\nc}{\newcommand}
\nc{\non}{\nonumber}
\nc{\hc}{\hbox {H.c.}} 
\nc{\noi}{\noindent}
\nc{\barx}{\bar{x}}
\nc{\pbarn}{\;\hbox {pb}}
\nc{\fbarn}{\;\hbox {fb}}
\nc{\lsp}{\;\;\;\;\;}
\nc{\Lsp}{\;\;\;\;\;\;\;\;\;\;}  
\nc{\LLsp}{\lspace \lspace}
\nc{\lra}{\longrightarrow}
\nc{\beq}{\begin{equation}}  \nc{\eeq}{\end{equation}}
\nc{\bea}{\begin{eqnarray}}  \nc{\eea}{\end{eqnarray}}
\nc{\baa}{\begin{array}}     \nc{\eaa}{\end{array}}
\nc{\bit}{\begin{itemize}}   \nc{\eit}{\end{itemize}}
\nc{\ben}{\begin{enumerate}} \nc{\een}{\end{enumerate}}
\nc{\bce}{\begin{center}}    \nc{\ece}{\end{center}}
\nc{\bpm}{\begin{pmatrix}}   \nc{\epm}{\end{pmatrix}}
\nc{\bvt}{\begin{verbatim}}  \nc{\evt}{\end{verbatim}}

\def\gesim{\,{\raise-3pt\hbox{$\sim$}}\!\!\!\!\!{\raise2pt\hbox{$>$}}\,}
\def\lesim{\,{\raise-3pt\hbox{$\sim$}}\!\!\!\!\!{\raise2pt\hbox{$<$}}\,}

\def\gev{\;\hbox{GeV}}
\def\tev{\;\hbox{TeV}}

\def\lsp{\qquad}
\def\lsim{\lesim}
\def\gsim{\gesim}
\def\hc{\hbox{H.c.}}
\def\vev{vacuum expectation value}
\def\lcal{{\cal L}}
\def\ocal{{\cal O}}

\def\zBB{{\mathbbm Z}}

\nc{\lam}{\lambda}
\nc{\Lam}{\Lambda}
\nc{\Lams}{\Lambda^2}
\nc{\mws}{m_W^2}
\nc{\mzs}{m_Z^2}
\nc{\mts}{m_t^2}
\nc{\mh}{m_h}
\nc{\mhs}{m_h^2}
\nc{\mvp}{m_\vp}
\nc{\mvps}{m_\vp^2}
\nc{\mw}{m_W}
\nc{\mz}{m_Z}
\nc{\mt}{m_t}
\nc{\mH}{m_{H^\pm}}
\nc{\ma}{m_A}
\nc{\ms}{m_S}
\nc{\vp}{\varphi}
\nc{\mpl}{m_\text{Pl}}
\nc{\sbb}{s_\beta}
\nc{\cbb}{c_\beta}
\nc{\sba}{s_{\beta-\alpha}}
\nc{\cba}{c_{\beta-\alpha}}
\nc{\stb}{s_{2\beta}}
\nc{\ctb}{c_{2\beta}}
\nc{\mb}{m_b}
\nc{\mbs}{m_b^2}
\nc{\tgb}{\tan\beta}
\nc{\tgbs}{\tan^2\beta}
\nc{\ctbs}{\cot^2\beta}

\nc{\lamp}{\lambda_H}
\nc{\lamvp}{\lambda_\varphi}
\nc{\lamx}{\lambda_x}
\nc{\xf}{x_f}

\nc{\co}{{\bf ??? }}

\renewcommand{\Re}{{\rm Re\thinspace}}
\renewcommand{\Im}{{\rm Im\thinspace}}

\usepackage[dvips]{color}
\definecolor{Black}{named}{Black}
\definecolor{Red}{named}{Red}

\allowdisplaybreaks

\begin{document}

\preprint{IFT-10-?? \cr }

\title{Tempered Two-Higgs-Doublet Model}

%
\author{B. Grzadkowski}
\email[]{Bohdan.Grzadkowski@fuw.edu.pl}
\affiliation{Institute of Theoretical Physics, University of Warsaw, 
Ho\.za 69, PL-00-681 Warsaw, Poland}
\author{P. Osland}
\email[]{per.osland@ift.uib.no}
\affiliation{Department of Physics and Technology, University of Bergen,
Postboks 7803, N-5020 Bergen, Norway}

\date{\today}

\begin{abstract}
  We discuss phenomenological consequences of requiring the
  cancellation of quadratic divergences up to the leading two-loop order
  within the Two-Higgs-Doublet Model (2HDM).  
  Taking into account existing experimental constraints, allowed
  regions in the parameter space, permitting the cancellation,
  are determined. A degeneracy between masses of scalar bosons is
  observed for $\tgb \gsim 40$. The possibility for CP violation in
  the scalar potential is discussed and regions of $\tgb-M_{H^\pm}$
  with substantial amount of CP violation are determined. In order to
  provide a source for dark matter in a minimal manner, a scalar gauge
  singlet is introduced and discussed. The model allows to ameliorate
  the little hierarchy problem by lifting the minimal scalar Higgs
  boson mass and by suppressing the quadratic corrections to
  scalar masses. The cutoff originating from the naturality arguments
  is therefore lifted from $\sim 0.6\tev$ in the Standard Model 
  to $\gsim 2.5 \tev$ in 2HDM, depending on the mass of the
  lightest scalar.

\end{abstract}


\maketitle

\section{Introduction}

The goal of this work is to extend the Standard Model (SM) such that 
there would be no quadratic divergences to scalar masses
up to the leading order at the two-loop level of the perturbation expansion.  
The quadratic divergences were first studied within the SM by
Veltman~\cite{Veltman:1980mj}, who showed that applying dimensional
reduction~\cite{Siegel:1979wq} one gets the following quadratically divergent
one-loop correction to the Higgs boson ($h$) mass 
\begin{equation}
\delta^\text{(SM)} \mhs = \frac{\Lams}{\pi^2 v^2}\left[\frac32
\mts-\frac18\left(6\mws+3\mzs\right) - \frac38 \mhs \right],
\label{hcor}
\end{equation}
where $ \Lam$ is a UV cutoff 
and $v \simeq 246 \gev $ denotes the \vev\ of the scalar doublet. The
issue of quadratic divergences was then investigated further adopting
other regularization schemes (e.g.\ point
splitting~\cite{Osland:1992ay}) and also in
\cite{Einhorn:1992um} without reference to any regularization scheme.

Since precision measurements require a light Higgs boson the
correction (\ref{hcor}) exceeds the mass itself even for small values
of $ \Lam $, e.g. for $\mh = 130 \gev$ we obtain $\delta^\text{(SM)}
\mh^2 \simeq \mh^2$ already for $\Lam \simeq 580 \gev$. On the other
hand, if we assume that the scale of new physics is widely separated
from the electro-weak scale, then constraints that emerge from
analysis of operators of dimension 6 require $\Lam \gsim$ a few TeV.
The lesson from this observation 
is that whatever is beyond the SM
physics, some amount of fine tuning is necessary; either we tune to
lift the cutoff above $\Lam \simeq 580 \gev$, or we tune when
precision observables measured at LEP are fitted.\footnote{In terms of the effective Lagrangian approach 
that implies coefficients of dimension-6 operators $c_i \ll 1$.}
Tuning both in corrections to the Higgs mass and in LEP physics is, of
course, also a viable alternative which we are going to explore below.
So, we will look for new physics in the TeV range which will allow to
lift the cutoff implied by quadratic corrections to $\mhs$ to the
multi-TeV range {\it and} which will be consistent with all the
experimental constraints---both require some amount of tuning.  
Within the SM the requirement $\delta^\text{(SM)} \mhs = 0$
implies $\mh \simeq 310~\gev$. However, as is very well know, the present
data favor a light SM Higgs boson --- according to the PDG~\cite{PDG}, after including all the
available experimental data and taking into account theoretical uncertainties,
the $99\%$ CL upper limit for the Higgs mass reads: $\mh \leq 194\gev$.
Therefore, within the SM the one-loop condition $\delta^\text{(SM)} \mhs = 0$
requires an unrealistic value of the Higgs boson mass.

Examining closer the experimental constraints one finds also the
following tension which emerges in the process of fitting all the
available data to the SM (see \cite{Chanowitz:2008ix} for a recent
review). Hadronic asymmetry measurements ($A^c_\text{FB},
A^b_\text{FB}, Q_\text{FB}$) favour a heavy Higgs boson, with $\mh
\sim 500 \gev$, while leptonic asymmetries ($A_\text{LR},
A^l_\text{FB}$) together with non-asymmetry precision measurements
($m_W, \Gamma_Z, \dots$) favour a Higgs mass smaller by one order of
magnitude. If ($A^c_\text{FB}, A^b_\text{FB}, Q_\text{FB}$)
are omitted from the fit one obtains $\mh\sim 50\gev$ with an
upper limit $\mh < 105\gev$ at the
95\%~CL~\cite{Chanowitz:2008ix}. Moreover there is the LEP lower limit
on the Higgs mass, $\mh > 114.4\gev$ \cite{Barate:2003sz}. The fit
which combines all the data is therefore of low quality. That
observation suggests a modification of the SM which would allow for a
heavy Higgs boson with a mass at least above the LEP limit.  For that
the SM prediction for the oblique parameters $S$ and $T$ must be
modified by the extension of the SM that we are seeking.

Here we are going to construct a model which would both soften the
little hierarchy problem by suppressing $\delta^\text{(SM)} \mhs$ {\it
  and} which would allow to lift the central value for the Higgs mass
up to a value which is well above the LEP limit (presumably it would
imply a better fit of the precision observables). We would like to point
out also that increasing the Higgs boson mass would ameliorate the
little hierarchy problem even if $\delta^\text{(SM)} \mhs$ was not
suppressed (since then larger cutoff would lead to the correction of
the order of the mass itself).
 
There are only two ways to suppress $\delta^\text{(SM)} \mhs/\mhs$:
one can either modify the SM such that {\it(i)} larger SM-like Higgs
boson mass is allowed; or {\it(ii)} extra radiative corrections to
$\delta^\text{(SM)} \mhs$ emerge that partially cancel
(\ref{hcor}). The best-studied example of the second approach is
provided by supersymmetric theories for which $\delta \mhs \ll \mhs$
up to the GUT scale, however the suppression of $\delta^\text{(SM)}
\mhs$ could also be achieved through very modest means, e.g.\ by
introducing just extra real scalar singlets to the
SM~\cite{Grzadkowski:2009mj} (although more tuning than in the
supersymmetric case is necessary). The first strategy was followed 
in \cite{Barbieri:2006dq} within the so-called
inert doublet model\footnote{The model was introduced in
  \cite{Deshpande:1977rw} in the context of dark matter.} (IDM).
There, a second Higgs doublet was introduced and an exact $\zBB_2$ 
symmetry was imposed to provide a dark-matter candidate.  
As shown in 
\cite{Barbieri:2006dq} a large ($400-600\gev$) SM-like Higgs boson 
mass was allowed by
the addition of an extra Higgs doublet (the inert doublet). It was
demonstrated that the extra contributions to the oblique parameters
originating from the inert doublet (with physical fields $H^\pm$, $A$
and $S$) can cancel large effects of a heavy SM Higgs, such that
$\mh\sim 400-600\gev$ is allowed. Here we propose a model which does
both, i.e. suppression of $\delta^\text{(SM)} \mh^2 $ by contributions
from some extra states (which implies reduced $\delta^\text{(SM)}
\mh^2/\mhs $) {\it and} which modifies the results of the global fit
such that much heavier SM Higgs boson is allowed (that also helps to
decrease $\delta^\text{(SM)} \mh^2/\mhs $ and in addition it could eliminate
the tension caused by the high LEP lower bound in the presence of a
low central value from precision tests).
A different approach to this problem has been proposed in \cite{Aoki:2008av}.

Another well-known problem of the SM is the strength of CP violation
(CPV) which is too weak to make the electroweak baryogenesis
viable \cite{Bernreuther:2002uj}. It is also worth noting a too slow
phase transition (within the SM) which is another difficulty for
realistic baryogenesis \cite{Bernreuther:2002uj}.

In light of the above remarks it seems very natural to consider simple
extensions of the SM scalar sector, as for instance the
Two-Higgs-Doublet Model (2HDM).  
Our intention is to bring the reader's attention to a region of parameter space
that not only is consistent with standard theoretical requirements (positivity
and unitarity) and satisfies all the relevant experimental constraints, but
also offers a simple pragmatic option to reduce the size of the quadratic corrections 
to scalar two-point Green's functions (so in other words to scalar masses).
It has been noticed a long time ago
\cite{Newton:1993xc} that within the 2HDM one can cancel quadratically
divergent corrections to two-point Green's functions for scalar
particles. Some phenomenological consequences of the cancellation were
discussed already in \cite{Ma:2001sj}.  
It is well known
\cite{Gunion:1989we} that within 2HDM extensions the oblique
parameters $S,T$ and $U$ can be modified such that SM contributions
growing with $\mh$ ($\propto \ln \mh$) could be canceled by other
terms (originating from extra scalars present in the 2HDM), so that
the lightest Higgs boson could be relatively heavy, see
\cite{Barbieri:2006dq}. Within the 2HDM the electroweak phase
transition could also be made fast enough~\cite{Fromme:2006cm} to make
electroweak baryogenesis viable.  The 2HDM provides also new sources
of CP violation in interactions of neutral scalars.  Therefore, here
we will discuss 2HDMs which do not suffer from quadratic divergences
in scalar two-point Green's functions, the tempered Two-Higgs-Doublet 
Model, seeking a model which also
allows for CP violation in the scalar potential. Since we have argued
above that the heavy SM-like Higgs boson would be more consistent with
experimental data, {\it we will investigate how much CP violation in
  2HDMs is allowed after lifting the Higgs mass well above the LEP
  lower limit and by imposing the conditions needed to cancel
  quadratic divergences (so as to ameliorate the little hierarchy
  problem)}.  We will not address here the issue of the electroweak
phase transition.
 
In a recent publication~\cite{Grzadkowski:2009bt}, motivated by
similar arguments, we have considered a version of the IDM with CP
violation introduced by replacing the SM-like Higgs doublet by a pair
of doublets. There, a candidate for dark matter (DM) was provided by
the lightest neutral component of the inert doublet (as in the
original IDM). In the model considered here, CP violation again
originates from the 2HDM, however in order to accommodate a DM candidate
{\it in a minimal manner} (instead of introducing the inert doublet as
in \cite{Grzadkowski:2009bt}) we extend the model by a real
singlet.\footnote{Although our basic motivations is different, this
  possibility is similar to the idea proposed in
  \cite{Kadastik:2009dj} for DM.} In fact, it is intriguing to note
that the singlet is even more inert than the original inert doublet
since it interacts only with the Higgs doublets and with right-handed
neutrinos, having no gauge interactions.

The paper is organized as follows. 
In Sec.~(\ref{non-IDM}) we investigate 
theoretical and phenomenological consequences of the cancellation
conditions within the general 2HDM. In order to accommodate a DM
candidate we introduce an extra real scalar gauge singlet, that is
discussed in Sec.~\ref{singlet-2hdm}. Section~\ref{sum} contains our
summary.

\section{ Non-inert Two-Higgs-Doublet Model}
\label{non-IDM}

A very appealing possibility would be to combine the IDM with the idea
of canceling the one-loop quadratic divergences. However, as we have shown 
in \cite{Grzadkowski:2009iz}, that is impossible because ot the vacuum stability 
conditions in the IDM are inconsistent with the requirement of cancellation of quadratic divergences. Since our intention is to build a 2HDM, which has no quadratic divergences at least at the one-loop level therefore, in the following 
we will consider a general (non-inert) 2HDM hoping for both a successful implementation 
of the cancellation condition and for a new source of CP violation. 
The price to pay will be the loss of a DM candidate. We return to that issue in Sec.~\ref{singlet-2hdm}.

In order to accommodate CP violation we consider here a non-inert 2HDM
with softly broken $\zBB_2$ symmetry which acts as $\Phi_1\to -\Phi_1$ and $u_R\to -u_R$ (all other fields are neutral). The scalar potential then reads
\begin{eqnarray}
V(\phi_1,\phi_2) &=&  -\frac12 \left\{m_{11}^2\phi_1^\dagger\phi_1 
+ m_{22}^2\phi_2^\dagger\phi_2 + \left[m_{12}^2 \phi_1^\dagger\phi_2 
+ \hc \right]\right\}
\label{2HDMpot} \\
&&  + \frac12 \lam_1 (\phi_1^\dagger\phi_1)^2 
+ \frac12 \lam_2 (\phi_2^\dagger\phi_2)^2 
+ \lambda_3(\phi_1^\dagger\phi_1)(\phi_2^\dagger\phi_2) 
+ \lambda_4(\phi_1^\dagger\phi_2)(\phi_2^\dagger\phi_1) 
+ \frac12\left[\lambda_5(\phi_1^\dagger\phi_2)^2 + \hc\right] \nonumber 
\end{eqnarray}
The minimization conditions at $\langle \phi_1^0 \rangle = v_1/\sqrt{2}$ and 
$\langle \phi_2^0 \rangle = v_2/\sqrt{2}$ can be formulated as follows:
\begin{eqnarray}
m_{11}^2&=&v_1^2\lam_1+v_2^2(\lambda_{345}-2\nu), \nonumber \\
m_{22}^2&=&v_2^2\lam_2+v_1^2(\lambda_{345}-2\nu),
\label{min}
\end{eqnarray}
where $\lambda_{345}\equiv \lam_3+\lam_4+\Re\lam_5$ and 
$\nu\equiv \Re m_{12}^2/(2v_1v_2)$.

We assume that $\phi_1$ and $\phi_2$ couple to down- and up-type
quarks, respectively (the so-called 2HDM II).  

\subsection{One-loop quadratic divergences}
\label{one-loop}

The cancellation of one-loop quadratic divergences for the scalar two-point Green's functions 
at zero external momenta ($G_i$, $i=1,2$) implies~\cite{Newton:1993xc} in the case of 2HDM
type II:
\begin{eqnarray}
G_1\equiv \frac32 \mw^2 + \frac34 \mz^2 
+ \frac{v^2}{2}\left( \frac32 \lam_1 + \lam_3 + \frac12 \lam_4 \right) 
- 3 \frac{\mb^2}{\cbb^2} = 0,
\label{qdcon1_mod2}\\
G_2\equiv\frac32 \mw^2 + \frac34 \mz^2 
+ \frac{v^2}{2}\left( \frac32 \lam_2 + \lam_3 + \frac12 \lam_4 \right) 
-3 \frac{\mt^2}{\sbb^2} = 0,
\label{qdcon2_mod2}
\end{eqnarray} 
where $v^2\equiv v_1^2+v_2^2$, $\tan\beta\equiv v_2/v_1$ and we adopt the
notation: $s_\beta \equiv \sin\beta$ and $c_\beta\equiv \cos\beta$.  We
note that when $\tan\beta$ is large, the two quark contributions can be
comparable. In the type II model
the mixed, $\phi_1-\phi_2$, Green's function is not quadratically divergent.

In the general CP-violating case, the quartic couplings $\lambda_i$ can be
expressed in terms of the mass parameters and elements of the rotation matrix 
needed for diagonalization of the scalar masses (see, for
example, Eqs.~(3.1)--(3.5) of \cite{ElKaffas:2007rq}):
\begin{align} \label{Eq:lambda1}
\lambda_1&=\frac{1}{c_\beta^2v^2}
[c_1^2c_2^2M_1^2
+(c_1s_2s_3+s_1c_3)^2M_2^2 \nonumber \\
&+(c_1s_2c_3-s_1s_3)^2M_3^2
-s_\beta^2\mu^2], \\
\lambda_2&=\frac{1}{s_\beta^2v^2} \label{Eq:lambda2}
[s_1^2c_2^2M_1^2
+(c_1c_3-s_1s_2s_3)^2M_2^2 \nonumber \\
&+(c_1s_3+s_1s_2c_3)^2M_3^2-c_\beta^2\mu^2], \\
\lambda_3&=\frac{1}{c_\beta s_\beta v^2} \label{Eq:lambda3}
\{c_1s_1[c_2^2M_1^2+(s_2^2s_3^2-c_3^2)M_2^2 \nonumber \\
&+(s_2^2c_3^2-s_3^2)M_3^2]
+s_2c_3s_3(c_1^2-s_1^2)(M_3^2-M_2^2)\} 
+\frac{1}{v^2}[2M_{H^\pm}^2-\mu^2], \\
\lambda_4&=\frac{1}{v^2}
[s_2^2M_1^2+c_2^2s_3^2M_2^2+c_2^2c_3^2M_3^2
+\mu^2-2M_{H^\pm}^2], \label{Eq:lambda4} \\
\Re\lambda_5&=\frac{1}{v^2}
[-s_2^2M_1^2-c_2^2s_3^2M_2^2-c_2^2c_3^2M_3^2+\mu^2], \\
\Im\lambda_5&=\frac{-1}{c_\beta s_\beta v^2}
\{c_\beta[c_1c_2s_2M_1^2-c_2s_3(c_1s_2s_3+s_1c_3)M_2^2 \nonumber \\
&+c_2c_3(s_1s_3-c_1s_2c_3)M_3^2] 
+s_\beta[s_1c_2s_2M_1^2 \label{Eq:lambda5I} \\
&+c_2s_3(c_1c_3\!-\!s_1s_2s_3)M_2^2
\!-\!c_2c_3(c_1s_3\!+\!s_1s_2c_3)M_3^2]\}, \nonumber 
\end{align}
where $\mu \equiv v^2 \nu$ while
$c_i=\cos\alpha_i$ and $s_i=\sin\alpha_i$ refer to the
neutral-Higgs-sector rotation matrix $R$, the latter parametrized
in terms of the angles $\alpha_1,\alpha_2$ and $\alpha_3$
according to the convention of \cite{Accomando:2006ga}.

It will be useful to adopt the following relation (emerging from the 
diagonalization of the neutral Higgs mass matrix~\cite{Khater:2003wq})
between $M_{1}^2$, $M_{2}^2$ and $M_3^2$:
\begin{equation}
M_3^2=\frac{M_1^2 R_{13} (-R_{11} + R_{12} \tgb)
+ M_2^2 R_{23} (-R_{21} + R_{22} \tgb)}{R_{33} (R_{31} - R_{32} \tgb)}.
\label{m3}
\end{equation}
Substitution of (\ref{m3}) into (\ref{Eq:lambda1})--(\ref{Eq:lambda5I})
allows to express the quartic couplings through the mixing angles
together with $M_{1}^2$, $M_{2}^2$, $M_{H^\pm}^2$ and $\mu^2$
(eliminating $M_3^2$).  Then, inserting the appropriate quartic
couplings into the conditions for cancellation of the quadratic
divergences (\ref{qdcon1_mod2})--(\ref{qdcon2_mod2}) we obtain two
linear equations for $M_{1}^2$ and $M_{2}^2$ with coefficients
depending on the mixing angles $\alpha_i$ (as well as on $M_{H^\pm}^2$
and $\mu^2$).  Therefore, for a given choice of $\alpha_i$'s, the
squared neutral-Higgs masses $M_{1}^2$, $M_{2}^2$ and $M_3^2$ can be
determined from the cancellation conditions
(\ref{qdcon1_mod2})--(\ref{qdcon2_mod2}) in terms of $\tgb$, $\mu^2$
and $M_{H^\pm}^2$.

\begin{figure}[ht]
\centering
\includegraphics[width=16cm]{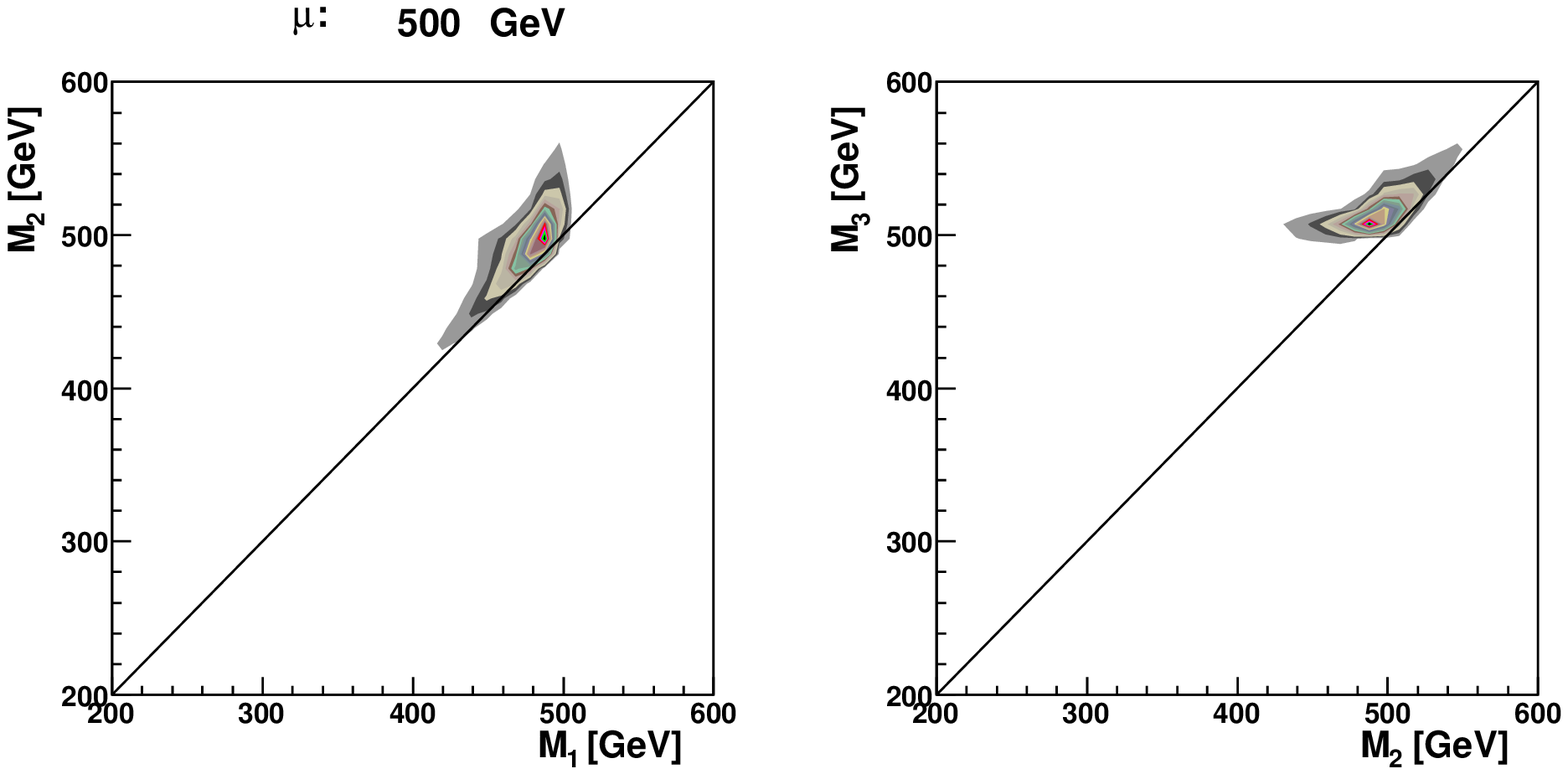}
\includegraphics[width=16cm]{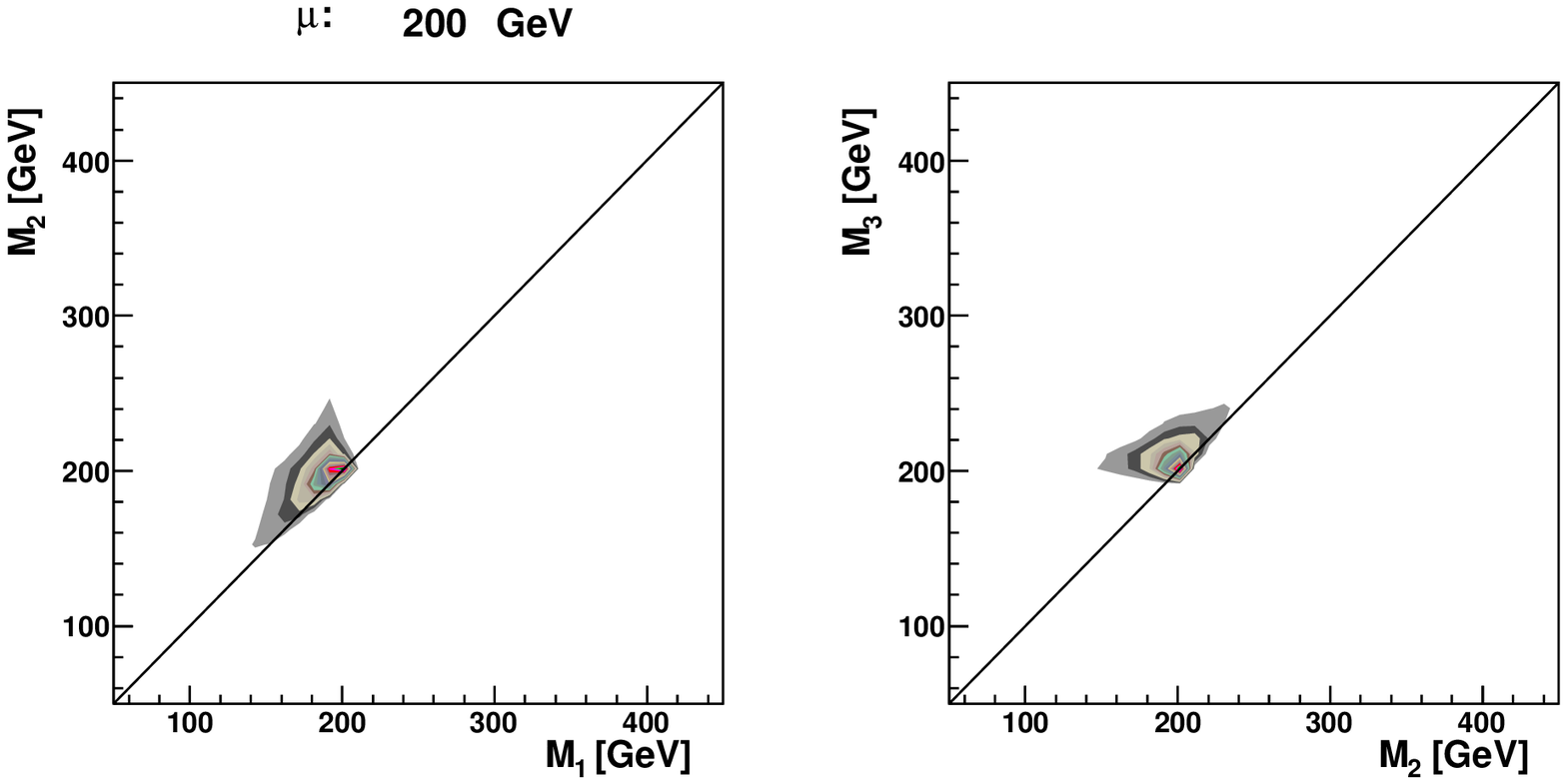}
\caption{
Distributions of allowed masses $M_2$ vs $M_1$ (left panels)
  and $M_3$ vs $M_2$ (right), resulting from a scan over the full
  range of $\alpha_i$, $\tan\beta \in (40,50)$ and $M_{H^\pm} \in
  (300,700)\gev$, for $\mu=500~\text{GeV}$ (top) and 
  $\mu=200~\text{GeV}$ (bottom).  No constraints are
  imposed other than the cancellation of quadratic divergences
  (\ref{qdcon1_mod2})--(\ref{qdcon2_mod2}), $M_i^2>0$ and
  $M_1<M_2<M_3$. The color coding indicates increasing
  density (while scanning over the parameter space)
  of allowed points as one moves inward from the boundary.}
\label{masses-500-0-0-0}
\end{figure}

Scalar masses resulting from a scan over $\alpha_i$, $M_{H^\pm}$ and 
$\tgb$ are shown in
Fig.~\ref{masses-500-0-0-0} for $\mu=200\gev$ and $\mu=500\gev$.  The charged Higgs
boson mass was varied between 300~GeV and 700~GeV.  Since only large
$\tan\beta$ will turn out to be allowed we have chosen to display plots
with $40 \leq \tan\beta \leq 50$ in order to illustrate a specific
property of the scalar spectrum that is visible at large $\tan\beta$.
Under the scan, the $M_i^2$ were calculated along the lines described
above. The only extra constraints (the cancellation of quadratic
divergences was, of course, guaranteed implicitly by the construction)
imposed were $M_i^2 > 0$ and $M_1 \leq M_2 \leq M_3$.  A striking
degeneracy of the neutral-Higgs masses is observed for the case of
large $\tan\beta$.  This degeneracy can be understood by expanding
$M_i^2$ for large $\tgb$.  The cancellation conditions,
Eqs.~(\ref{qdcon1_mod2})--(\ref{qdcon2_mod2}) can then be expressed as
follows:
\begin{align}
Y_{11} M_1^2 + Y_{12} M_2^2 - Y_{13} (4\mbs + \mu^2) 
&= \ocal\left(\frac{1}{\tgb}\right), \label{con1exp}\\
2 R_{12}R_{22}\tgb (M_1^2-M_2^2) -
R_{33}[-4\bar{m}^2-2M_{H^\pm}^2+12\mts+\mu^2] 
&= \ocal\left(\frac{1}{\tgb}\right),
\label{con2exp}
\end{align}
where\footnote{Note that $Y_{11}+Y_{12}=Y_{13}$.}
\begin{align}
Y_{11}
&\equiv-R_{12}R_{13}R_{31}^2 + R_{11}^2 R_{32}R_{33}, \nonumber \\
Y_{12}
&\equiv-R_{22}R_{23}R_{31}^2 + R_{21}^2R_{32}R_{33}, \nonumber \\
Y_{13}
&\equiv R_{32}R_{33}, \nonumber \\
\bar m^2
&\equiv \frac32\mws + \frac34 \mzs.
\end{align}
First, it is useful to notice that (\ref{con2exp}) implies
\begin{equation}
M_1^2-M_2^2 
\sim \frac{1}{\tgb}\frac{R_{33}}{R_{12}R_{22}} 
[-4\bar{m}^2-2M_{H^\pm}^2+12\mts+\mu^2]
\end{equation}
Therefore, for $\tgb \gg 1$, we expect to have $M_1^2 \simeq M_2^2$
unless $|R_{12}R_{22}|\ll 1$ or $-2M_{H^\pm}^2+\mu^2$ is very
large.\footnote{Note that cancellations between the $M_{H^{\pm}}^2$ and
  $\mu^2$ terms are possible. This is why the degeneracy survives even for
  $\mu$ as large as $\mu=500\gev$, see Fig.~\ref{masses-500-0-0-0}.}
Secondly, solving (\ref{con1exp})--(\ref{con2exp}) one finds that to
leading order (for large $\tan\beta$) $M_1^2=M_2^2=\mu^2+4\mb^2$.

On the other hand, expanding (\ref{m3}) for $\tgb \gg 1$ one obtains:
\begin{equation}
M_3^2=-\frac{M_1^2 R_{12}R_{13}+M_2^2 R_{22} R_{23}}{R_{32} R_{33}} 
+ {\cal{O}}\left(\frac{1}{\tgb}\right).
\label{m3exp}
\end{equation}
Therefore (invoking unitarity of $R$) it is seen that the degeneracy
$M_1 = M_2$ implies that also $M_1 = M_2 = M_3$.  Finally we can
conclude that for large $\tgb$ one obtains $M_1 \simeq M_2 \simeq M_3
\simeq \mu^2+4\mb^2$, this explains the approximate degeneracy observed in
Fig.~\ref{masses-500-0-0-0}.\footnote{The
  reader should be warned that the above expansions are justified if
  the coefficients of sub-leading terms $\propto 1/\tan\beta$ are not
  enhanced by special values of the mixing angles (that would
  correspond to CP conservation in the scalar sector). Since here we
  are interested in the case of CP violation, we will not elaborate on
  those CP conserving limits.}

\subsection{Two-loop leading quadratic divergences}
\label{two-loop}

The generic form of the quadratically divergent contributions to scalar 
two-point Green's functions at zero external momenta reads~\cite{Einhorn:1992um}
\begin{equation} 
\delta G_i = \Lam^2
\sum_{n=0}f_n^{(i)}(\lam)
\left[ \ln\left(\frac{\Lam}{\bar\mu}\right) \right]^n + \cdots \,,
\label{quad_cor}
\end{equation} 
where $n$ corresponds to $(n+1)$-loop contribution, $\lam$
stands for relevant coupling constants, $\bar\mu$ is the renormalization scale
and $f_n^{(i)}(\lam)$ is a calculable
(order by order) function (polynomial) of the couplings. It should be noticed
that at the $(n+1)$-loop level there exist also sub-leading contributions that contain terms $\propto \Lam^2(\ln \Lam)^m$ with $m<n$, so for instance at the two-loop level the leading contribution
is $\propto \Lam^2 \ln \Lam$ while there are also sub-leading terms $\propto \Lam^2$.
The coefficients of the leading terms,
$f_n^{(i)}(\lam)$, can be determined recursively adopting a nice algorithm noticed by Einhorn and Jones~\cite{Einhorn:1992um}:
\beq
(n+1)f_{n+1}^{(i)}=\bar\mu\frac{\partial}{\partial \bar\mu}f_n^{(i)}=
\sum_I \beta_{\lambda_I}\frac{\partial}{\partial \lambda_I}f_n^{(i)}
\label{recur}
\eeq
where the sum runs over coupling constants that contribute to the coefficient $f_n^{(i)}$. Hereafter we will limit ourselves to the leading two-loop contributions. 
Therefore, to calculate $f_1^{(i)}$, only the one-loop coefficient $f_0^{(i)}$ and
one-loop beta functions are needed. As beta functions for the 2HDM are known~\cite{Kominis:1993zc} the cancellation condition for quadratic divergences up to the leading two-loop order can easily be determined:
\beq
G_1+\delta G_1=0 \lsp {\rm and} \lsp G_2+\delta G_2=0 
\label{2-loop-con}
\eeq
with
\bea
\delta G_1 &=& \frac{v^2}{8} [
9 g_2 \beta_{g_2} + 3 g_1 \beta_{g_1} + 6\beta_{\lambda_1} + 4 \beta_{\lambda_3} + 2 \beta_{\lambda_4}]\ln\left(\frac{\Lambda}{\bar\mu}\right)\\
\delta G_2 &=& \frac{v^2}{8} [
9 g_2 \beta_{g_2} + 3 g_1 \beta_{g_1} + 6\beta_{\lambda_2} + 4 \beta_{\lambda_3} + 2 \beta_{\lambda_4}
-24 g_t \beta_{g_t}]\ln\left(\frac{\Lambda}{\bar\mu}\right)
\eea
In what follows, adopting (\ref{Eq:lambda1})--(\ref{Eq:lambda5I}) we will be solving the conditions (\ref{2-loop-con}) for the scalar masses $M_i^2$ for a given set of $\alpha_i$'s, $\tgb$, $\mu^2$ and $M_{H^\pm}^2$. For the renormalization scale
we will adopt $v$, so $\bar\mu=v$.\footnote{Since we are using tree-level relations between quartic couplings and scalar masses, the renormalization scale should be of the order of the masses themselves, that is why we adopt here $\bar \mu=v$. For a more exhaustive discussion of the renormalization-scale dependence, see the first paper of \cite{Kolda:2000wi}.} Then those masses together with the corresponding coupling constants, will be adopted to find predictions of the model for various observables which can be confronted with experiments.

\subsection{Positivity and unitarity constraints}
\label{posit-unit}

The requirements of positivity for the 2HDM model potential are well known
\cite{Deshpande:1977rw}:
\begin{gather}
\lam_{1,2} > 0, \label{stab12} \\
\lam_3 > - \sqrt{\lam_1 \lam_2}, \quad
\lam_L\equiv\lam_3+\lam_4-|\lam_5| > - \sqrt{\lam_1 \lam_2}.
\label{stab345}
\end{gather} 
One could also require the above conditions to be satisfied up the
unification scale $\Lambda \sim 10^{15}\gev$. That approach, resulting in much stronger constraints
in terms of allowed scalar masses and $\tan\beta$  was followed
in Ref.~\cite{Ferreira:2009jb}. Here, as we consider UV completion
appearing at the scale of a few TeV we do not follow that line of reasoning.

\subsection{Experimental constraints}
\label{exp}

We impose the following experimental constraints:
\bit
\item The oblique parameters $T$ and $S$
\item $B_0-\bar{B}_0$ mixing
\item $B\to X_s \gamma$
\item $B\to \tau \bar\nu_\tau X$
\item $B\to D\tau \bar\nu_\tau$
\item LEP2 Higgs-boson non-discovery
\item $R_b$
\item The muon anomalous magnetic moment
\item Electron electric dipole moment 
\eit 
For more details concerning the implementation of the experimental constraints, see
refs.~\cite{Grzadkowski:2009bt,ElKaffas:2007rq,WahabElKaffas:2007xd}.
Subject to all these constraints, we find allowed solutions of (\ref{2-loop-con}). 
The recent paper \cite{Deschamps:2009rh} also contains an exhaustive analysis of 
experimental constraints on the 2HDM type II. The lower limit on the
charged Higgs-boson mass adopted here, $M_{\mp}\geq 300\gev$ (basically
determined by the $b\to s \gamma$ constraint) agrees roughly with
the $95\%$ CL lower limit, $316\gev$, obtained in \cite{Deschamps:2009rh}
irrespectively of $\tan\beta$.

\begin{figure}[t]
\centering
\includegraphics[width=5.5cm]{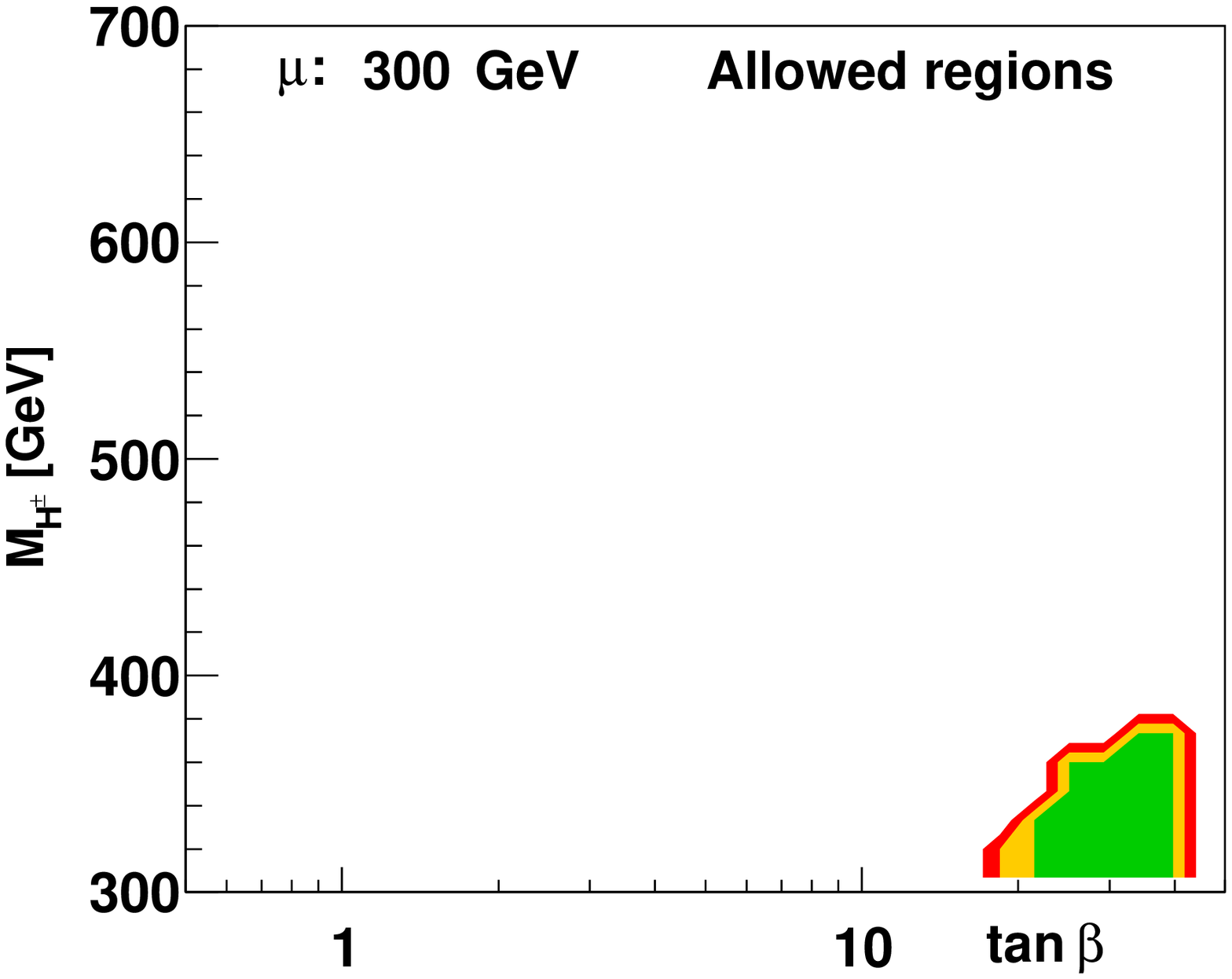}
\includegraphics[width=5.5cm]{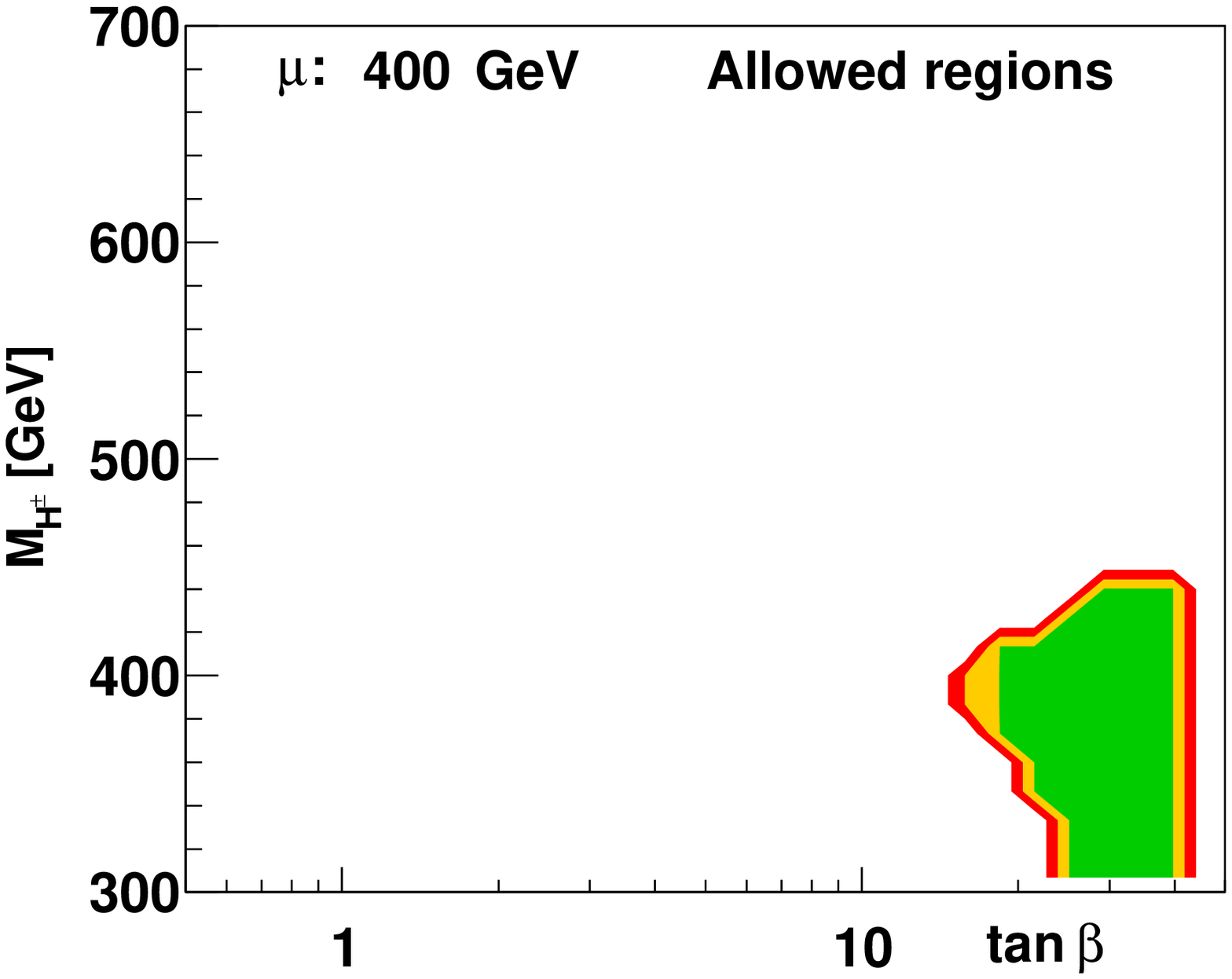}
\includegraphics[width=5.5cm]{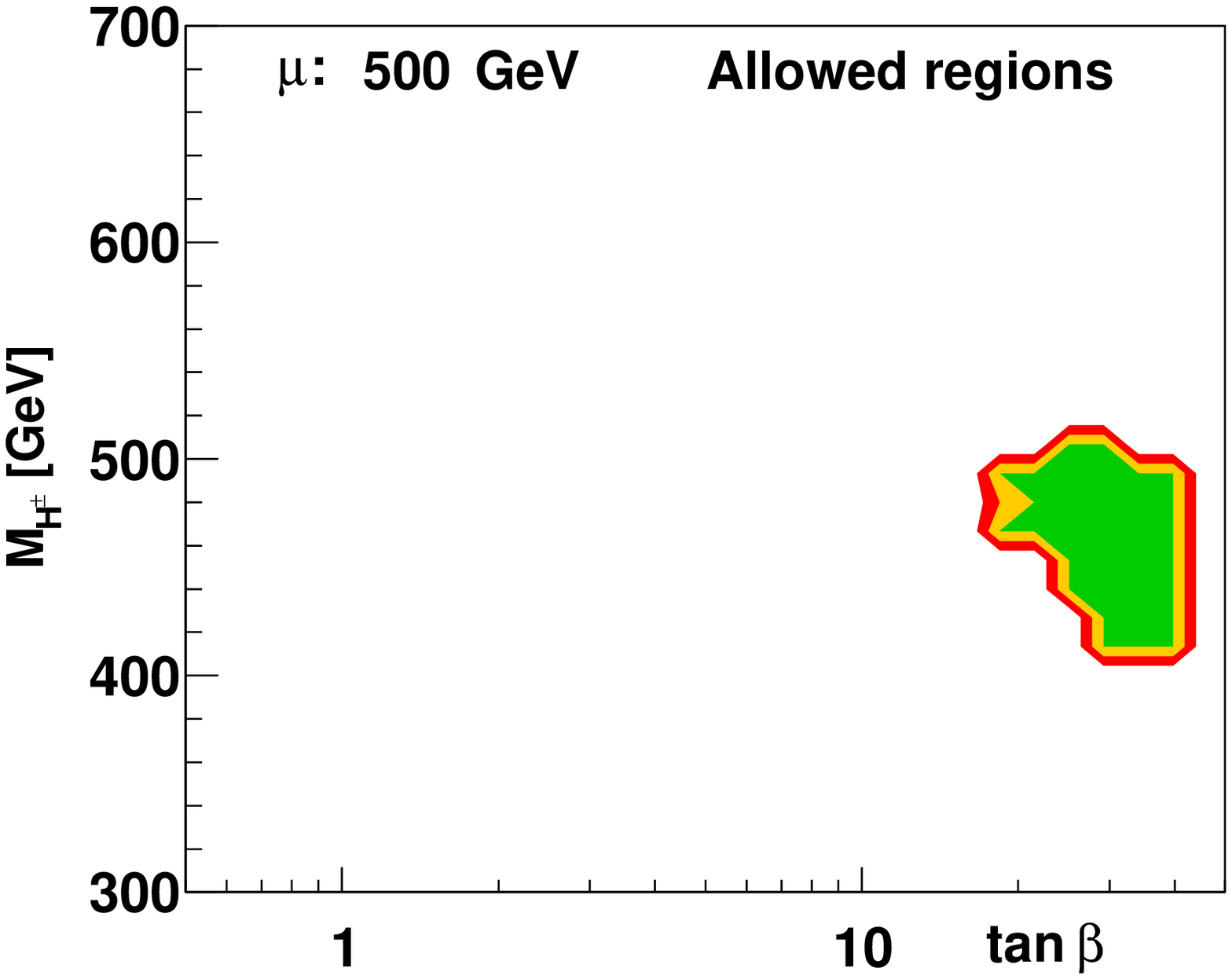}
\caption{\label{Fig:allowed-2500-300-400-500}
  Two-loop allowed regions in
  the $\tan\beta$--$M_{H^\pm}$ plane, for $\Lam=2.5\tev$, for $\mu=300, 400, 500\gev$ 
  (as indicated).  Red: positivity
  is satisfied; yellow: positivity and unitarity both satisfied;
  green: also experimental constraints satisfied at the 95\% C.L., as
  specified in the text. }
\end{figure}

\begin{figure}[t]
\centering
\includegraphics[width=5.5cm]{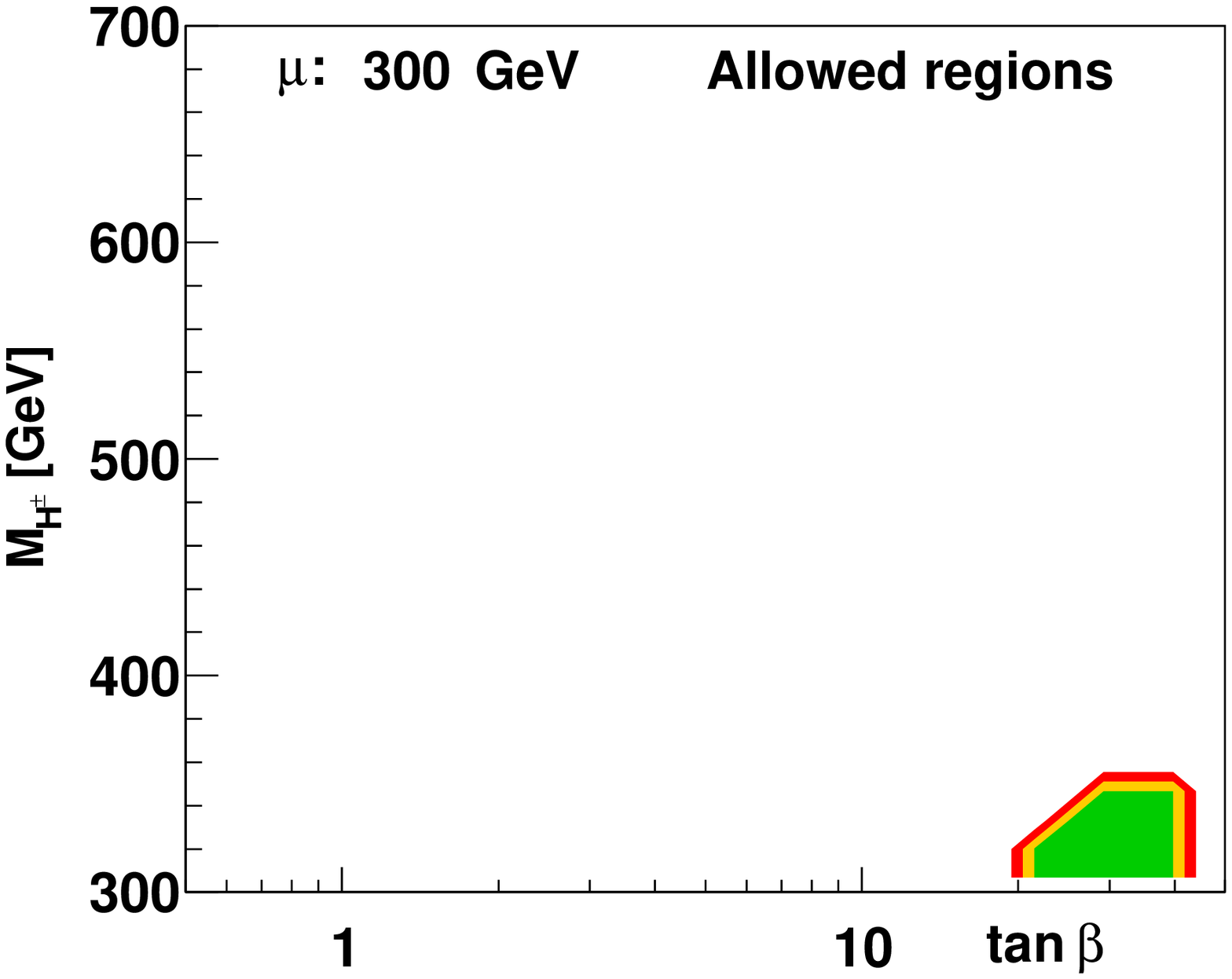}
\includegraphics[width=5.5cm]{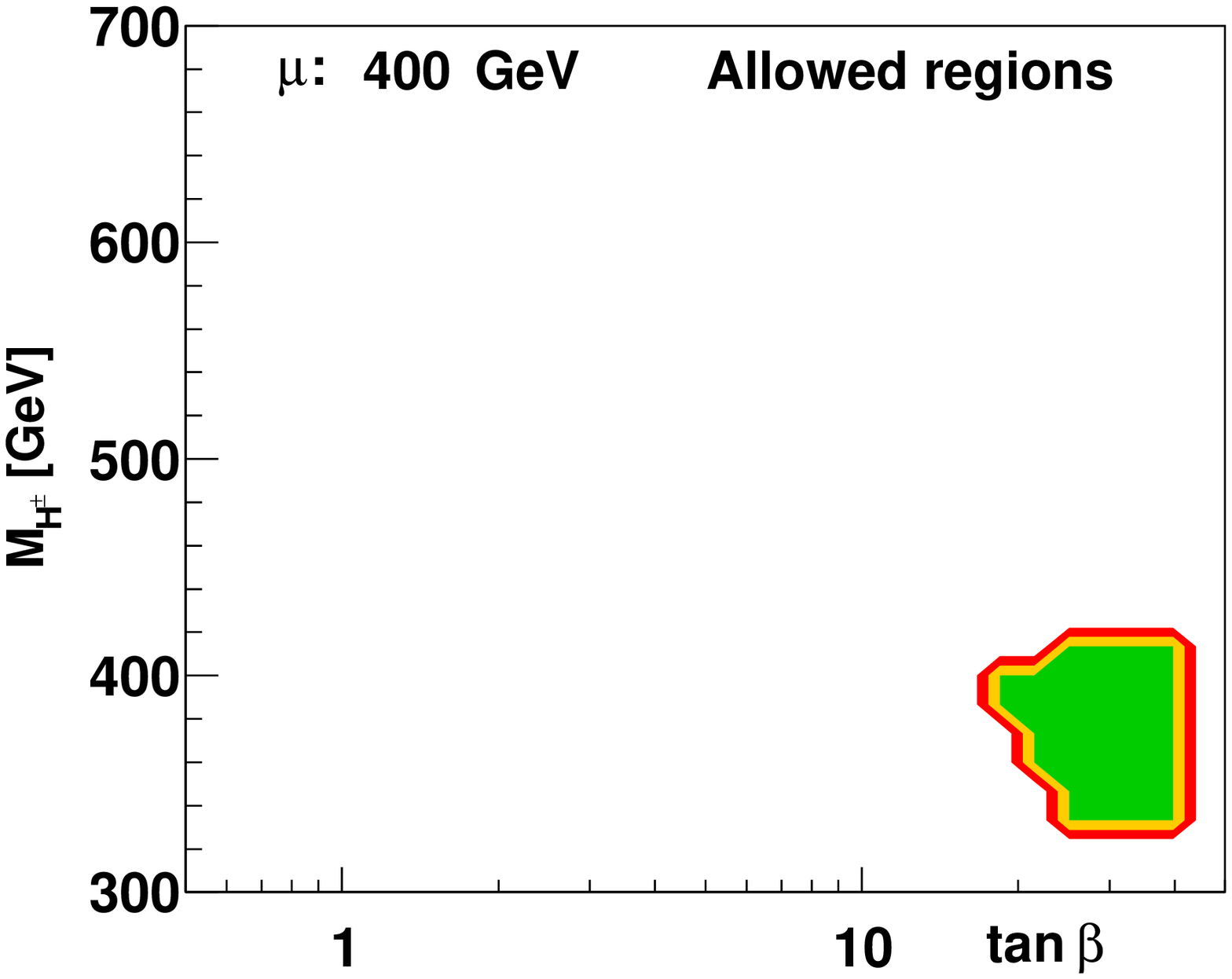}
\includegraphics[width=5.5cm]{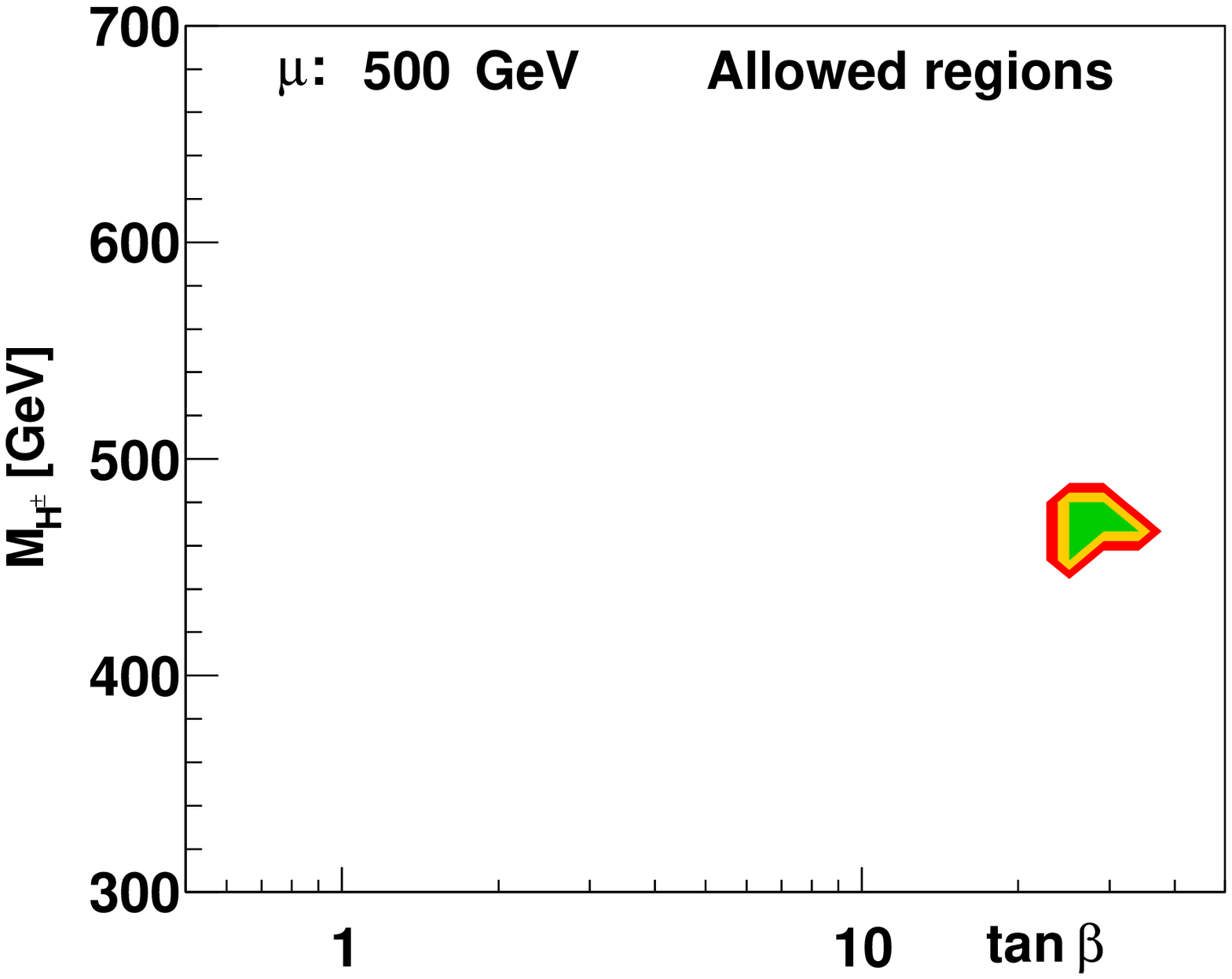}
\caption{\label{Fig:allowed-6500-300-400-500}
Similar to Fig.~\ref{Fig:allowed-2500-300-400-500} for 
$\Lam=6.5\tev$. 
}
\end{figure}

\subsection{Allowed regions}
\label{sec:allowed}

Imposing the above conditions 
we find allowed regions in the $\tan\beta$--$M_{H^\pm}$ plane as 
illustrated by the red domains in the $\tan\beta$--$M_{H^\pm}$ plane, see
Figs.~\ref{Fig:allowed-2500-300-400-500} and \ref{Fig:allowed-6500-300-400-500}
for fixed values of $\mu$. The allowed regions
were obtained scanning over the mixing angles $\alpha_i$ and solving
the two-loop cancellation conditions (\ref{2-loop-con}). Imposing
also unitarity in the Higgs-Higgs-scattering sector 
\cite{Kanemura:1993hm,Akeroyd:2000wc,Ginzburg:2003fe}
(yellow regions), the allowed regions are only slightly
reduced. Requiring that also experimental constraints listed in
the Sec.~\ref{exp} are satisfied one obtains green regions shown
in the figure. 

\begin{figure}[ht]
\centering
\includegraphics[width=12cm]{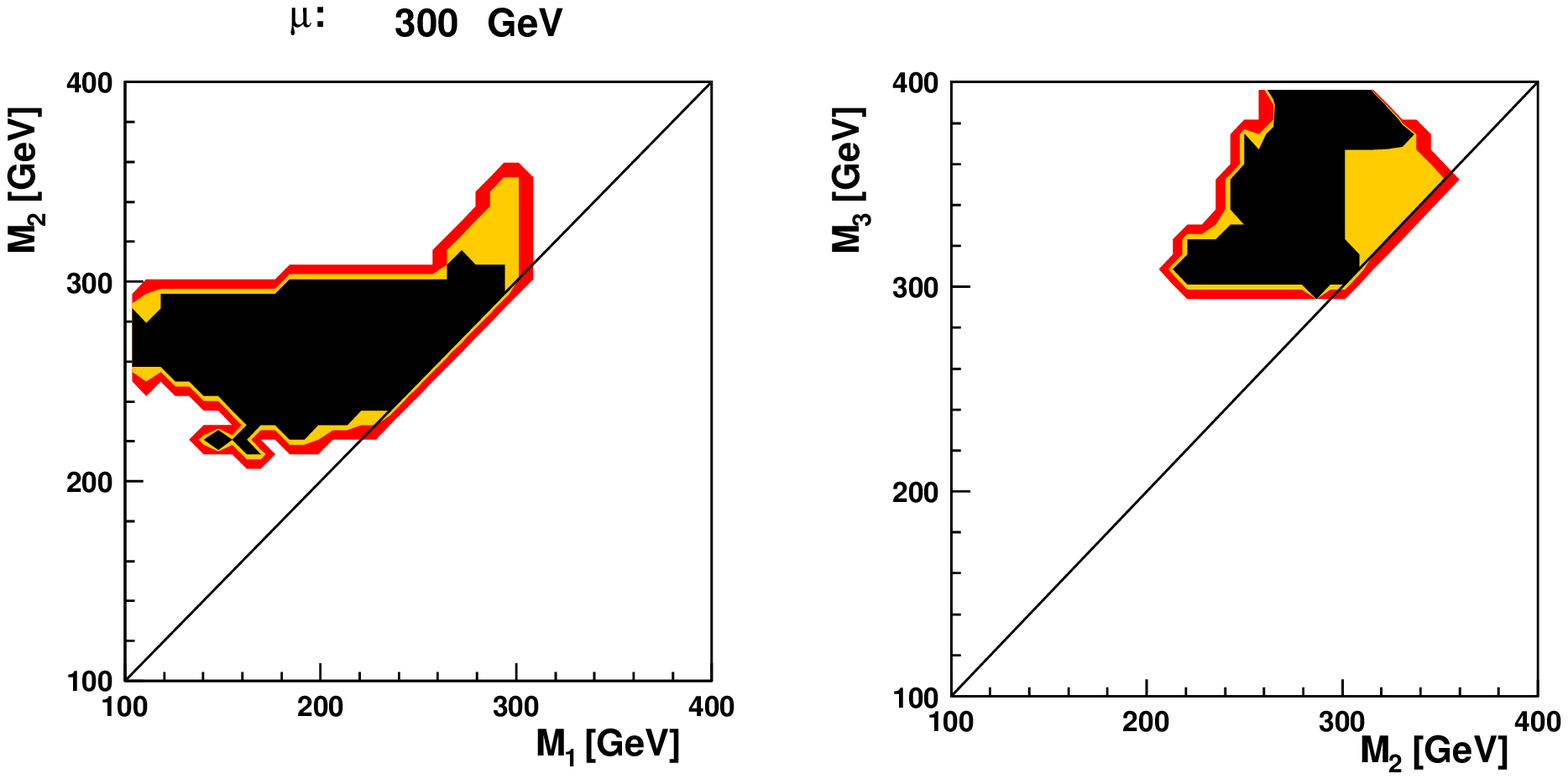}
\includegraphics[width=12cm]{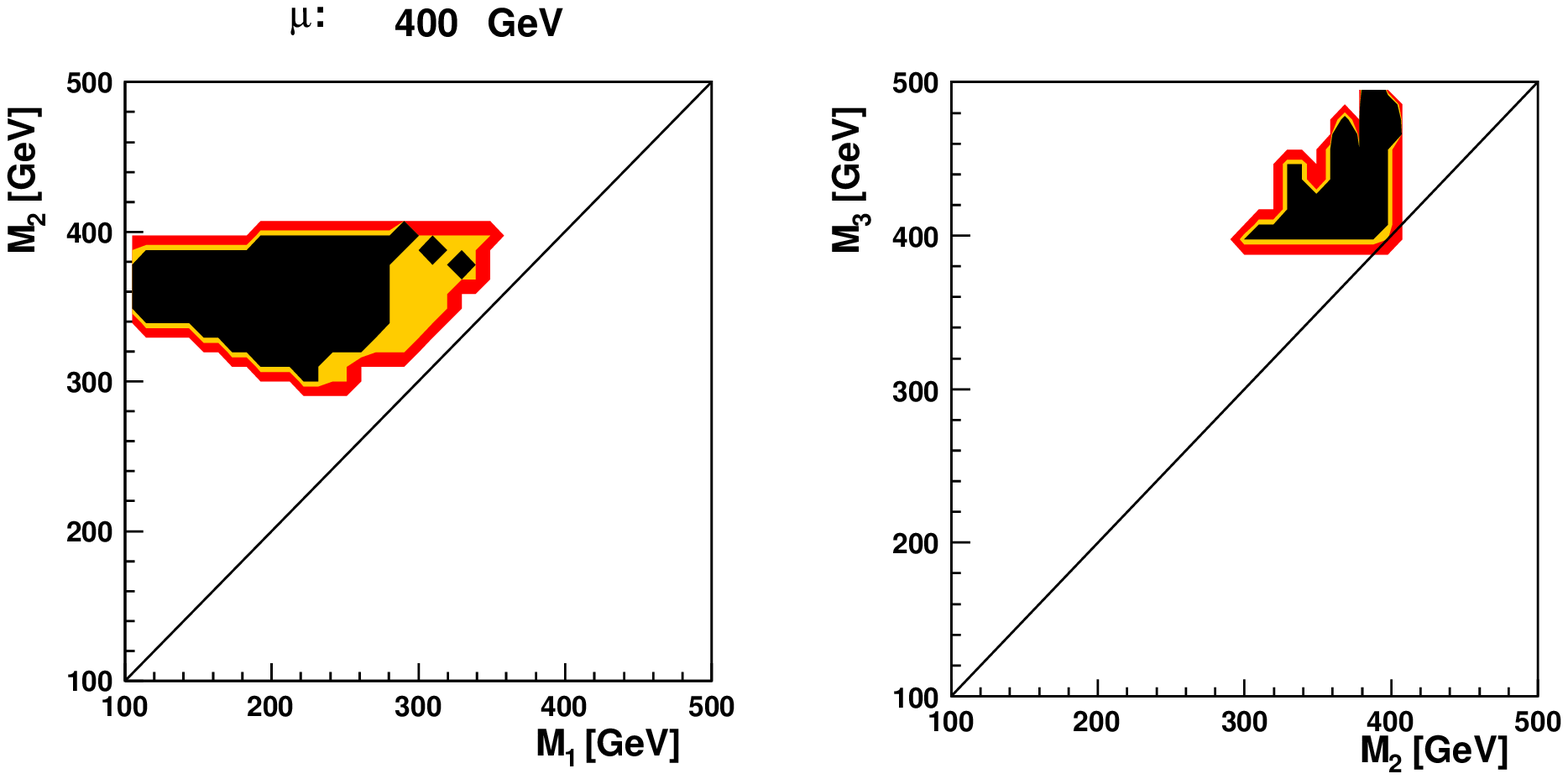}
\includegraphics[width=12cm]{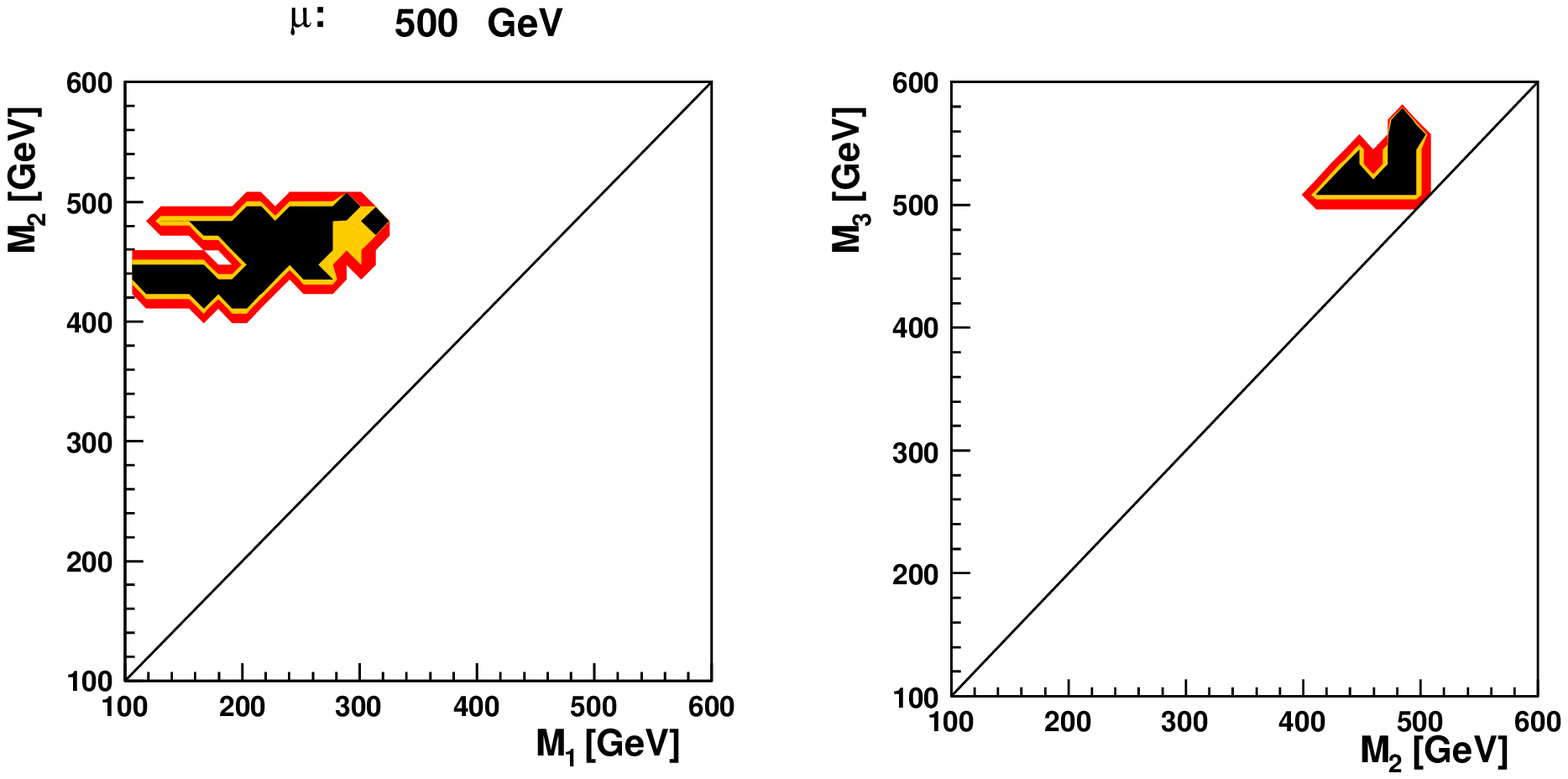}
\caption{
  Two-loop distributions of allowed masses $M_2$ vs $M_1$ (left panels)
  and $M_3$ vs $M_2$ (right) for $\Lam=2.5\tev$, resulting from a scan over the full
  range of $\alpha_i$, $\tan\beta \in (0.5,50)$ and $M_{H^\pm} \in
  (300,700)\gev$, for $\mu=300, 400, 500~\text{GeV}$. Red: Positivity
  is satisfied; yellow: positivity and unitarity both satisfied;
  green: also experimental constraints satisfied at the 95\% C.L., as
  specified in the text. }
\label{Fig:2-loop-masses-2500}
\end{figure}

\begin{figure}[ht]
\centering
\includegraphics[width=12cm]{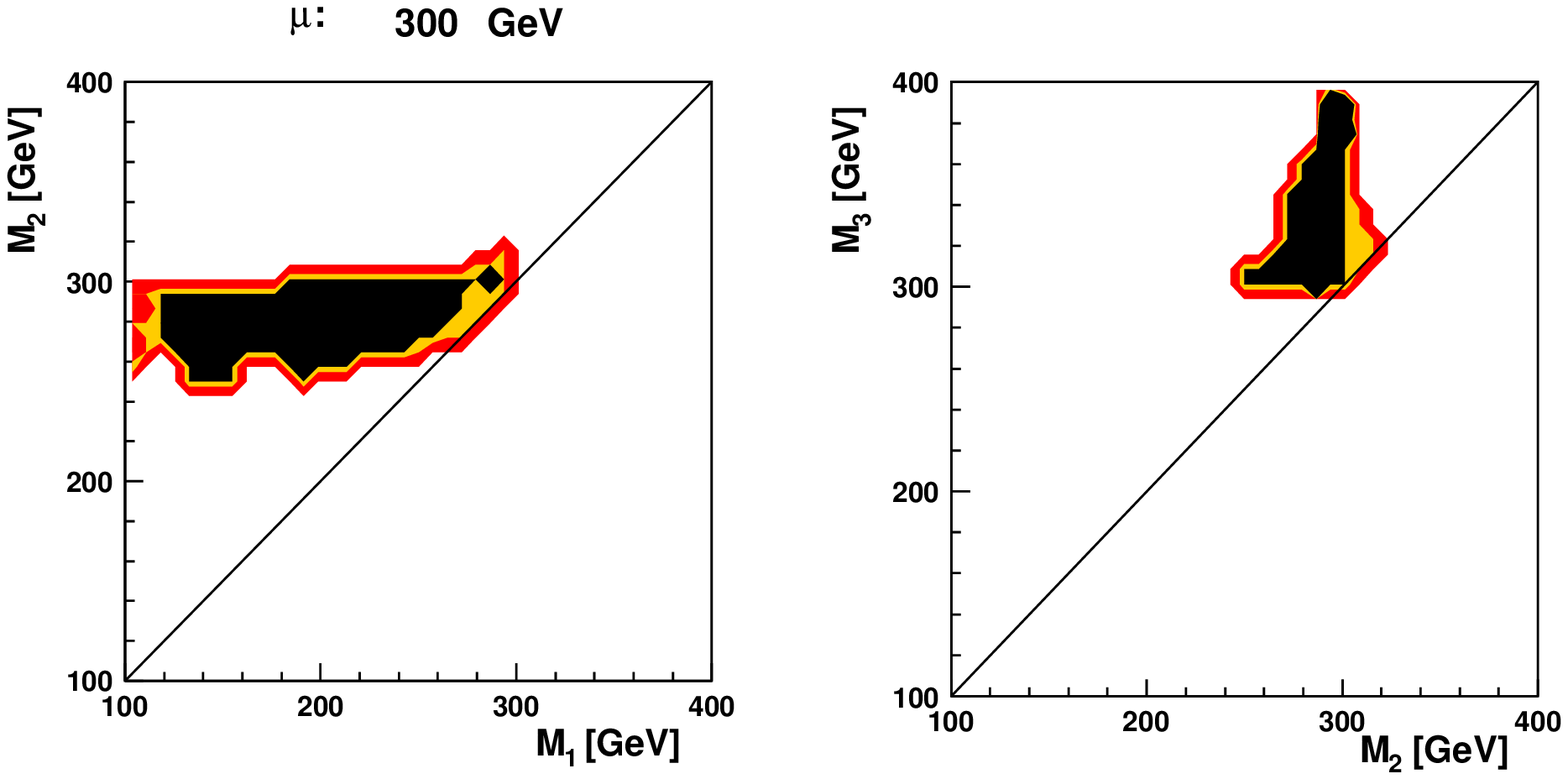}
\includegraphics[width=12cm]{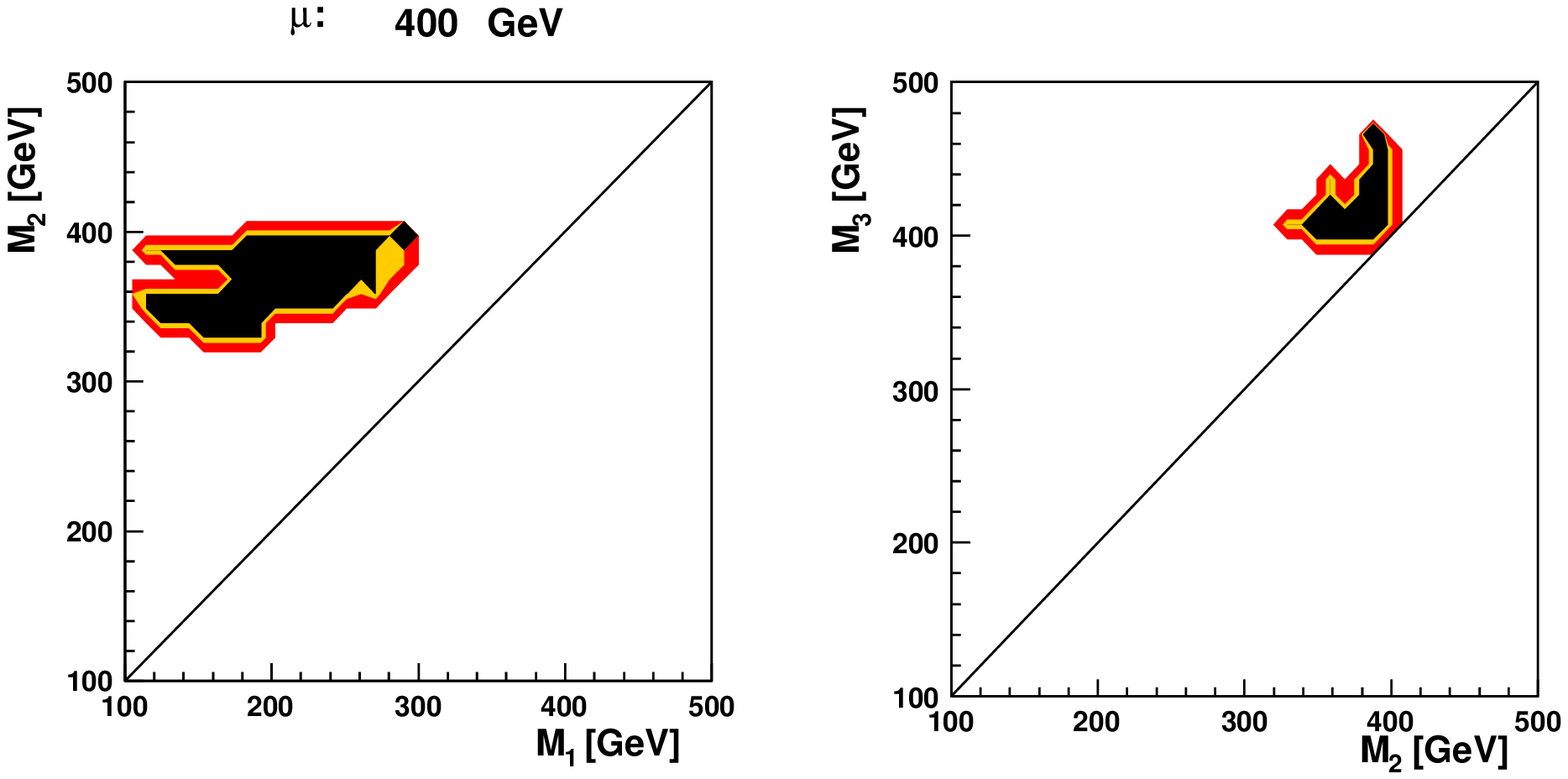}
\includegraphics[width=12cm]{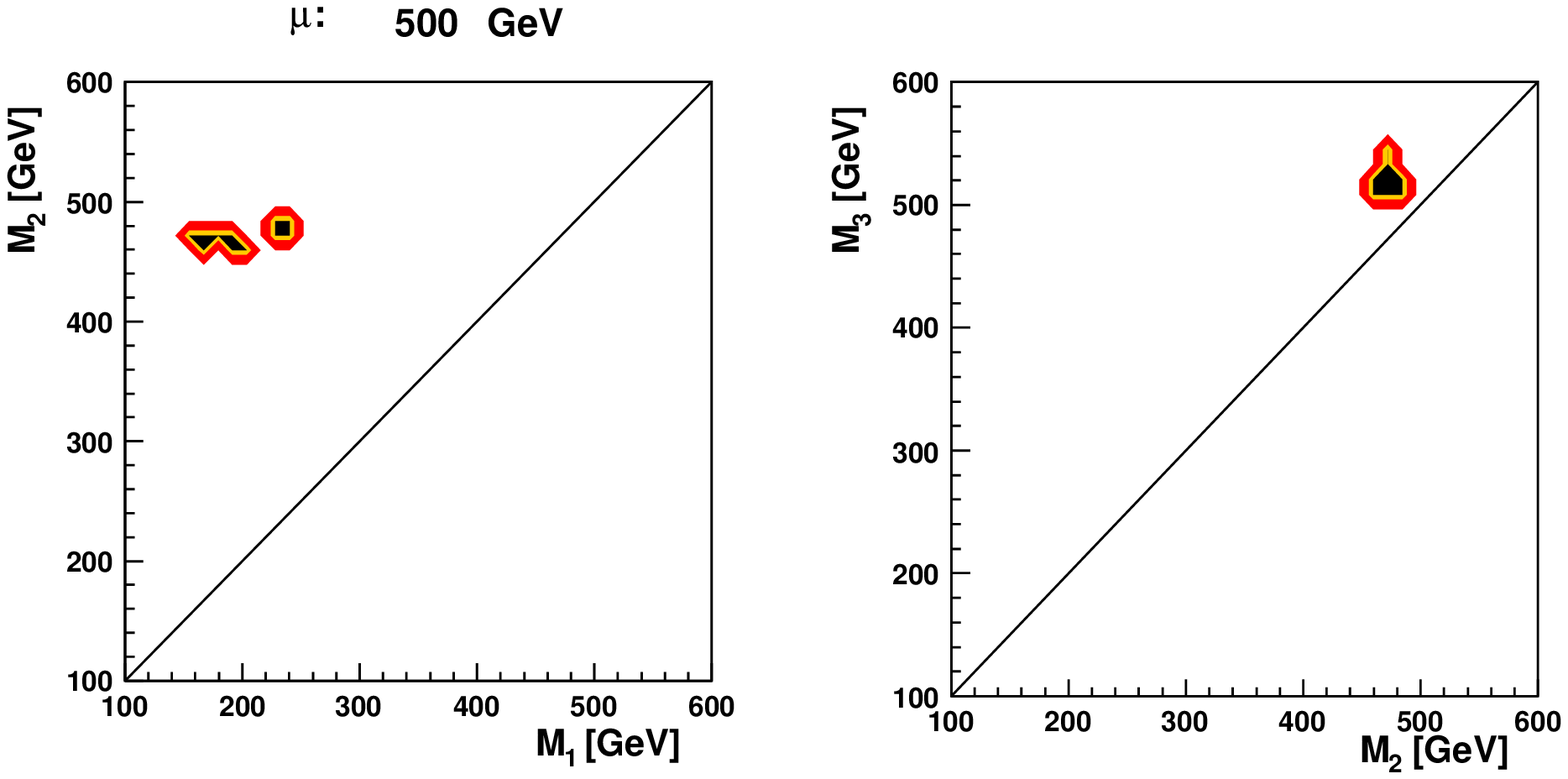}
\caption{Similar as Fig.~\ref{Fig:2-loop-masses-2500} for $\Lam=6.5\tev$
  for $\mu=300, 400, 500~\text{GeV}$. }
\label{Fig:2-loop-masses-6500}
\end{figure}

For parameters that are consistent with unitarity, positivity, experimental
constraints and the two-loop cancellation conditions (\ref{2-loop-con}), we show in Figs.~\ref{Fig:2-loop-masses-2500}-\ref{Fig:2-loop-masses-6500} scalar masses resulting from a scan over $\alpha_i$, 
$M_{H^\pm}$ and $\tgb$. Those plots could be compared with Fig.~\ref{masses-500-0-0-0}. One should however remember
that in the two-loop case also unitarity, positivity and experimental constraints
are taken into account.
Note that in Figs.~\ref{Fig:allowed-2500-300-400-500} and \ref{Fig:allowed-6500-300-400-500}, consistent solutions are obtained only for $\tan\beta \gsim 15$ while at the one-loop level, also a small low-$\tan\beta$ region was allowed after imposing all the constraints. That small low-$\tan\beta$ region is disallowed after the two-loop corrections are imposed, see \cite{Grzadkowski:2009iz} for the one-loop result.
As we have noticed for the one-loop spectrum, large $\tan\beta$ implies
similar scalar masses. This is indeed what is being observed in Figs.~\ref{Fig:2-loop-masses-2500}-\ref{Fig:2-loop-masses-6500}
also for the two-loop case. The allowed solutions ``peak'' around $M_{H^\pm}\sim \mu$ with $20 \lsim \tan\beta \lsim 50$. For $\mu=200$ and 600~GeV there are hardly any solutions for
$\Lam=2.5~\text{TeV}$ and no solutions were found for $\Lambda=6.5~\text{TeV}$.

\subsection{CP violation}
\label{cpv}

Here we are going to discuss the
possibility of having CP violation in the scalar potential
(\ref{2HDMpot}), subject to the two-loop cancellation of quadratic
divergences (\ref{2-loop-con}).  In
order to parametrize the magnitude of CP violation we adopt the rephasing
invariants introduced by Lavoura and Silva \cite{Lavoura:1994fv}
(see also \cite{Branco:2005em}). We shall here
use the basis-invariant
formulation of these invariants $J_1$, $J_2$ and $J_3$ as
proposed by Gunion and Haber \cite{Gunion:2005ja}.
As is proven there (theorem \#4) the Higgs sector is
CP-conserving if and only if all $J_i$ are real. In the basis adopted
here the invariants read~\cite{Grzadkowski:2009bt}:
\begin{eqnarray}
\Im J_1&=&-\frac{v_1^2v_2^2}{v^4}(\lambda_1-\lambda_2)\Im \lambda_5,
\label{Eq:ImJ_1} \\
\Im J_2&=&-\frac{v_1^2v_2^2}{v^8}
\left[\left((\lambda_1-\lambda_3-\lambda_4)^2-|\lambda_5|^2\right) v_1^4
+2(\lambda_1-\lambda_2) \Re \lambda_5 v_1^2v_2^2\right.\nonumber\\
&&\hspace*{1.2cm}\left.
-\left((\lambda_2-\lambda_3-\lambda_4)^2-|\lambda_5|^2\right) v_2^4\right]
\Im \lambda_5,\label{Eq:ImJ_2} \\
\Im J_3&=&\frac{v_1^2v_2^2}{v^4}(\lambda_1-\lambda_2)
(\lambda_1+\lambda_2+2\lambda_4)\Im \lambda_5.
\label{Eq:ImJ_3}
\end{eqnarray}
It is seen that there is no CP violation when $\Im\lambda_5=0$,
see \cite{Grzadkowski:2009bt} for more details.

\begin{figure}[ht]
\centering
\includegraphics[width=16cm]{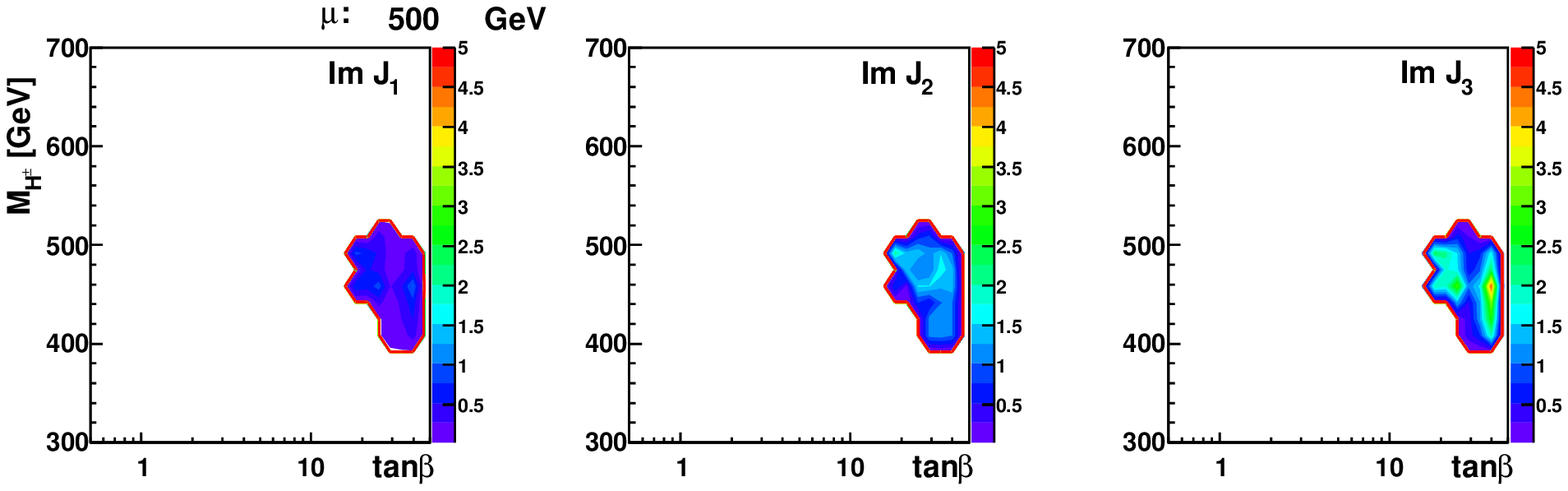}
\includegraphics[width=16cm]{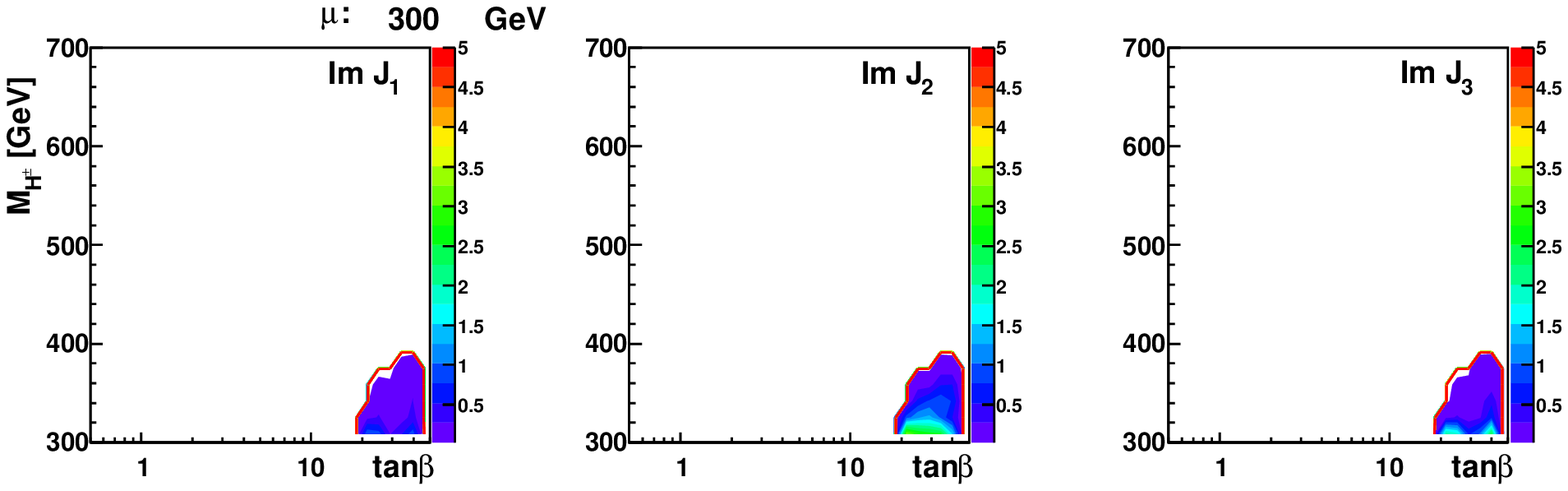}
\caption{
  Imaginary parts of the rephasing invariants $|\Im J_i|$ at the
  two-loop level for $\Lam=2.5\tev$, for
  $\mu=500~\text{GeV}$ (top) and $\mu=300~\text{GeV}$ (bottom). 
  The colour coding in units $10^{-3}$ is given along the
  right vertical axis.}
\label{Fig:imj-2500-300-500}
\end{figure}

\begin{figure}[ht]
\centering
\includegraphics[width=16cm]{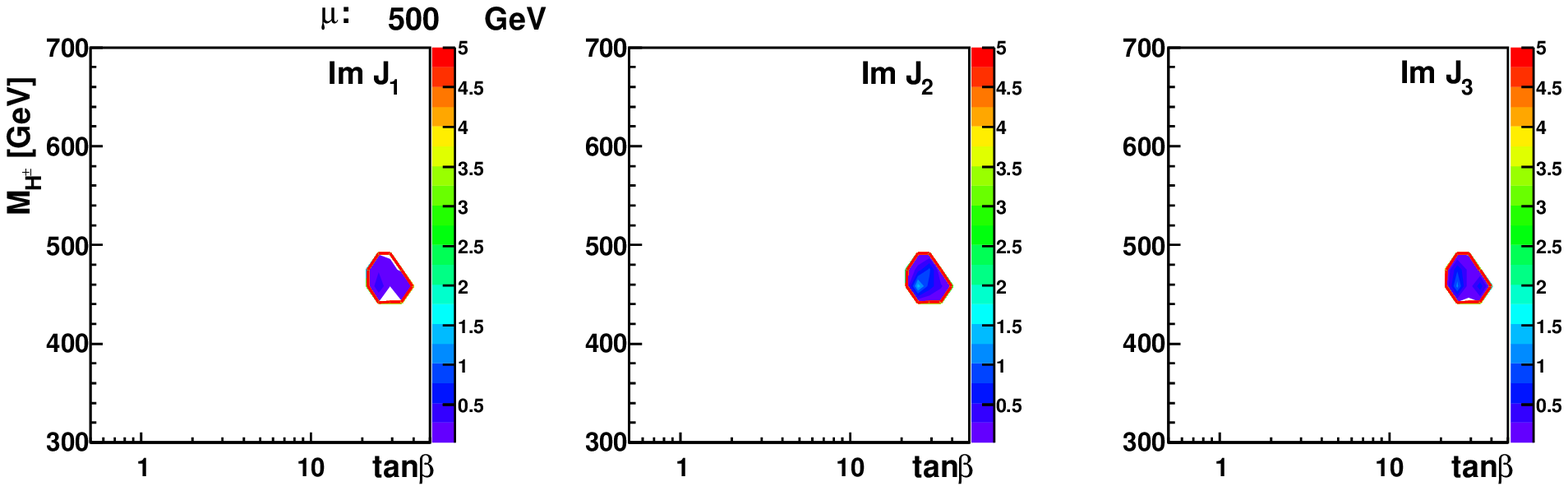}
\includegraphics[width=16cm]{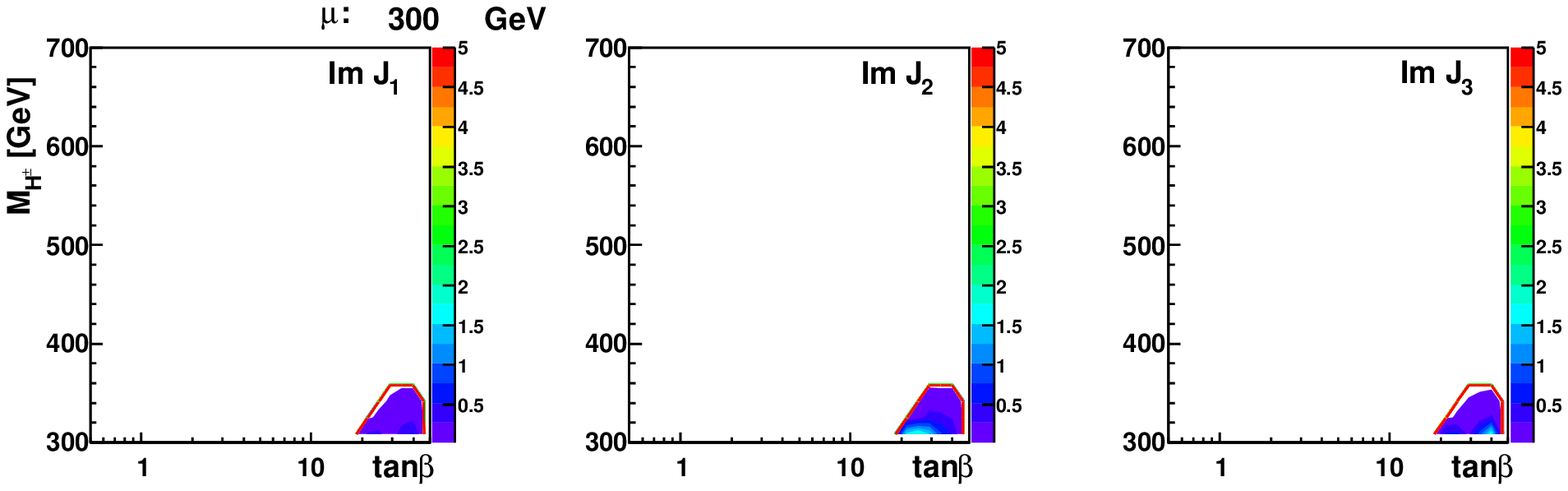}
\caption{Similar as Fig.~\ref{Fig:imj-2500-300-500} for $\Lam=6.5\tev$.}
\label{Fig:imj-6500-300-500}
\end{figure}

As we have noted earlier, $\tgb$ above $\sim 40$ implies approximate
degeneracy of scalar masses. That could be catastrophic for CP
violation since it is well known that the exact degeneracy $M_1=M_2=M_3$
results in vanishing invariants $\Im J_i$ and no CP violation (exact
degeneracy implies $\Im \lambda_5=0$). Using the one-loop conditions
(\ref{qdcon1_mod2})--(\ref{qdcon2_mod2}) one immediately finds that
$\lambda_1-\lambda_2=4(\mb^2/\cbb^2-\mt^2/\sbb^2)/v^2$, which implies
\begin{equation}
\Im J_1 = 4\, \Im \lambda_5 \frac{\cbb^2\mt^2-\sbb^2 \mb^2}{v^2}
=-4\, \Im \lambda_5 \left(\frac{\mb}{v}\right)^2 + 
{\cal{O}}\left(\frac{\Im \lambda_5}{\tgb^2}\right)
\label{imj1}
\end{equation}
In fact the above result shows even more than we have anticipated.  If
$\tgb$ is large then $\Im J_1$ is suppressed not only by $\Im
\lambda_5 \simeq 0$ (as caused by $M_1\simeq M_2\simeq M_3$) but also
by the factor $(\mb^2/v^2)$, as implied by the cancellation conditions
(\ref{qdcon1_mod2})--(\ref{qdcon2_mod2}). The same suppression factor
appears for $\Im J_3$. The case of $\Im J_2$ is more involved, however
when $\mb^2/v^2$ is neglected all the invariants
(\ref{Eq:ImJ_1})--(\ref{Eq:ImJ_3}) have the same simple asymptotic
behavior
\begin{equation}
\Im J_i \sim \frac{\Im \lambda_5}{\tgbs}
\end{equation}
for large $\tgb$.  It is also worth noticing that $\tgb=\mt/\mb (\simeq
38)$ implies $\lambda_1=\lambda_2$, which in turn leads to exact
vanishing of $\Im J_1$ and $\Im J_3$.  Qualitatively those conclusions
survive at the two-loop level. For a quantitative illustration
we plot in Figs.~\ref{Fig:imj-2500-300-500}-\ref{Fig:imj-6500-300-500} maximal values of the invariants in
the $\tan\beta$--$M_{H^\pm}$ plane with all the necessary constraints
imposed, seeking regions which still allow for substantial CP
violation. At high values of $\tan\beta$ these invariants are of the
order of $10^{-3}$, in qualitative agreement with the discussion
above. Note that the SM corresponding invariant ${\rm Im} Q = (V_{ud} V_{cb} V_{ub}^\star V_{cd}^\star) \simeq 2\times 10^{-5} \sin \delta_{KM}$~\cite{Bernreuther:2002uj}, for
$V_{ij}$ and $\delta_{KM}$ being elements of the CKM matrix and CP-violating phase, respectively. Therefore the model considered here offers at least two orders of magnitude enhancement comparing to the SM.

\subsection{Stability and the determination of the cutoff}
\label{stab}

It should be emphasized here that the conditions
(\ref{2-loop-con}) eliminate the quadratic
divergences only up to the leading two-loop corrections.
Even though the sub-leading two-loop and higher effects
are suppressed by powers of coupling constants and 
powers of $1/(16\pi^2)$, nevertheless since the $\ln\Lam$ term is growing, there exists always $\Lam$ large enough, that the
hierarchy problem reappears: loop corrections to masses are again of the
order of the masses itself. In fact that observation allows to determine
the value of the cutoff up to which higher order corrections do not reintroduce the hierarchy problem (see \cite{Kolda:2000wi} for the analogous strategy within the SM).  In general, quadratic corrections to scalar
masses have the form of (\ref{quad_cor} ) 
\begin{equation} 
\delta M_i^2 = \Lam^2
\sum_{n=0}f_n^{(i)}(\lam)
\left[ \ln\left(\frac{\Lam}{v}\right) \right]^n + \cdots\,,
\label{loopcor}
\end{equation} 
where $v$ is chosen as a renormalization scale.
The following na\"ive estimation of $f_n^{(i)}$ is sufficient:
\begin{equation}
f_n^{(i)} \sim  \left(\frac{4 \pi}{16 \pi^2}\right)^{n+1} = \left(\frac{1}{4 \pi}\right)^{n+1}
\label{fn}
\end{equation}
where the relevant coupling constants were conservatively assumed to
be of the order of $4\pi$.\footnote{This estimate agrees qualitatively
  with the two-loop result obtained for $f_1$ in the SM, see
  Eq.~(21) in \cite{Kolda:2000wi}.}  Here we choose as the cutoff the
maximal value of $\Lam$ such that the higher order corrections do not exceed the mass of the lightest scalar, $M_1$:
\begin{equation}
\Lam \lsim 4\pi M_1
\label{Lam}
\end{equation}
Then, e.g. for $M_1=200$ $(500)\gev$ the cutoff is at least at $\Lam\sim 2.5$ $(6.3)\tev$. Of course, larger $M_1$ would imply higher $\Lam$. 

Having the cutoff determined, we should address the issue of higher-loop corrections to the equations (\ref{2-loop-con})
that ensure vanishing of the quadratic corrections up to leading two-loop effects.
As is seen from (\ref{loopcor}), the generic form of the condition for vanishing quadratic divergence is the following
\begin{equation}
\frac{\lam}{(4 \pi)^2} + \frac{\lam^2}{(4\pi)^4} \ln\left(\frac{\Lam}{v}\right)
+ \frac{\lam^2}{(4\pi)^4} + \frac{\lam^3}{(4\pi)^6} \ln^2\left(\frac{\Lam}{v}\right) + \cdots = 0\,.
\label {cor}
\end{equation} 
where $\lam$ stands for a typical coupling constant. The last two terms shown above (sub-leading two- and leading three-loop effects) have been neglected in the present analysis. It is then easy to see that even for 
the cutoff as large as $\Lam=6.5\tev$ using a very conservative (large) value for the typical coupling, $\lam=4\pi$, the precision of the adopted approximation 
is of the order of $12\%$. Note that whenever $\lam < 4\pi$ or $\Lam < 6.5\tev$, the adopted approximations work better.

\section{2 Doublet + 1 Singlet Higgs Model: the case for dark matter }
\label{singlet-2hdm}

In this scenario we combine CP violation present in the non-inert 2HDM
(allowing for softly broken $\zBB_2$ symmetry) with a real scalar $\vp$
which is a gauge singlet.  The singlet provides a natural DM
candidate (see \cite{Silveira:1985rk}, \cite{Burgess:2000yq} and
\cite{Grzadkowski:2009mj}). In this case the scalar potential is the
following
\begin{eqnarray}
V(\phi_1,\phi_2) &=&  -\frac12 \left\{m_{11}^2\phi_1^\dagger\phi_1 
+ m_{22}^2\phi_2^\dagger\phi_2 + \left[m_{12}^2 \phi_1^\dagger\phi_2 
+ \hc \right]\right\} 
+ \frac12 \lam_1 (\phi_1^\dagger\phi_1)^2 
+ \frac12 \lam_2 (\phi_2^\dagger\phi_2)^2 
\nonumber
\\
&& 
+ \lambda_3(\phi_1^\dagger\phi_1)(\phi_2^\dagger\phi_2) 
+ \lambda_4(\phi_1^\dagger\phi_2)(\phi_2^\dagger\phi_1) 
+ \frac12\left[\lambda_5(\phi_1^\dagger\phi_2)^2 + \hc\right] \nonumber 
\\
&&  + \mu_\vp^2\vp^2 + \frac{1}{24}\lam_\vp \vp^4 
+ \vp^2(\eta_1 \phi_1^\dagger\phi_1 + \eta_2 \phi_2^\dagger\phi_2).
\label{2HDM1Spot}
\end{eqnarray}
Note that the term $\propto \vp^2 \phi_1^\dagger \phi_2$ is forbidden as
it breaks the $\zBB_2$ symmetry in a hard way.
Since $\vp$ is supposed to be the DM candidate, in order to ensure its 
stability we have imposed an extra discrete symmetry
$\zBB_2^\prime$ such that $\vp \to -\vp$ while other fields are neutral. 
The symmetry excludes terms
odd in $\vp$. The potential should be arranged such that
the symmetry remains unbroken, so that $\langle \varphi \rangle = 0$.
For that it is sufficient to require 
\begin{equation}
\mu_\vp^2>0 \And \lam_\vp,\eta_1,\eta_2>0
\label{mincon}
\end{equation}
Then it is easy to see that if the standard 2HDM stability conditions
(\ref{stab12})--(\ref{stab345}) are fulfilled then the potential
(\ref{2HDM1Spot}) is also positive definite. For the mass of the
singlet we obtain: $m_\vp^2=2\mu_\vp^2+\eta_1 v_1^2+\eta_2 v_2^2$.

Since $\vp$ is a gauge singlet, therefore in the presence of right-handed 
neutrinos which are also gauge singlets 
the following Yukawa interaction is allowed~\cite{Grzadkowski:2009mj}: 
\begin{equation}
\lcal_Y = - \vp \overline{(\nu_R)^c} Y_\vp \nu_R + \hc
\label{lyuk}
\end{equation} 
Note that for a number of right-handed neutrino flavours greater than $1$, 
the Yukawa matrix $Y_\vp$ is
in general (depending on the quantum numbers of $\nu_R$  under $\zBB_2^\prime$, 
see \cite{Grzadkowski:2009mj}) non-vanishing.

In this model the conditions for cancellation of quadratic divergences are 
slightly modified:
\begin{eqnarray}
\frac32 \mw^2 + \frac34 \mz^2 + \frac{v^2}{2}\left(\frac12 \eta_1 
+ \frac32 \lam_1 + \lam_3 + \frac12 \lam_4 \right) 
- 3 \frac{\mb^2}{\cbb^2} &=& 0, \nonumber \\
\frac32 \mw^2 + \frac34 \mz^2 + \frac{v^2}{2}\left(\frac12 \eta_2 
+ \frac32 \lam_2 + \lam_3 + \frac12 \lam_4 \right) 
- 3 \frac{\mt^2}{\sbb^2}&=& 0 , \nonumber \\
\frac{\lambda_\vp}{2} + 4 (\eta_1+\eta_2) 
- 8{\rm Tr} \{Y_\vp Y_\vp^\dagger\} &=& 0.
\label{qdcon_mod3}
\end{eqnarray}
The last condition above guarantees vanishing quadratic divergence 
in corrections to the $\vp$ mass. 
Since for the positivity we assumed $\lam_\vp,\eta_1,\eta_2 > 0$,
it is clear from the above equation that the presence of the
Yukawa coupling $Y_\vp$ is mandatory to extend the condition for
cancellation of quadratic divergences to the singlet field as well.
It should also be mentioned that the presence of the singlet does
influence the two-loop corrections to the quadratic divergences, 
those effects are neglected as being small ($\propto \eta_i$).

As we have already mentioned the extra singlet $\vp$ provides a candidate for
the DM.  
To estimate its present abundance we consider the dominant
annihilation channels for $\vp$.  
The Lagrangian describing relevant cubic and quartic scalar interactions reads
\begin{equation}
\lcal=-\vp^2 (\kappa_i v H_i + \lam_{ij} H_i H_j 
+ \lam_\pm H^+ H^-) ,
\label{lag}
\end{equation}
where
\begin{align}
\kappa_i &= \eta_1 R_{i1}\cbb + \eta_2 R_{i2}\sbb \label{kap},\\
\lam_{ij} &= \frac12\left[\eta_1(R_{i1} R_{j1}+\sbb^2 R_{i3} R_{j3})+
                           \eta_2(R_{i2} R_{j2}+\cbb^2 R_{i3} R_{j3})\right], 
\label{Eq:lambda_ij}\\
\lambda_\pm&=\eta_1 s_\beta^2+\eta_2 c_\beta^2.
\label{lam}
\end{align}

A detailed study of the DM within this model (with extended conditions
for the cancellation of quadratic divergences (\ref{qdcon_mod3})) will
be presented elsewhere~\cite{dm}. However here we would like to show
that it is indeed natural to expect the right DM abundance in the
presence of the singlet. For an illustration we will assume that the
DM annihilation cross section is of the order of the contributions from the
lightest neutral Higgs boson $H_1$ of mass $M_1$. 

\begin{figure}[ht]
\centering
\includegraphics[width=16cm]{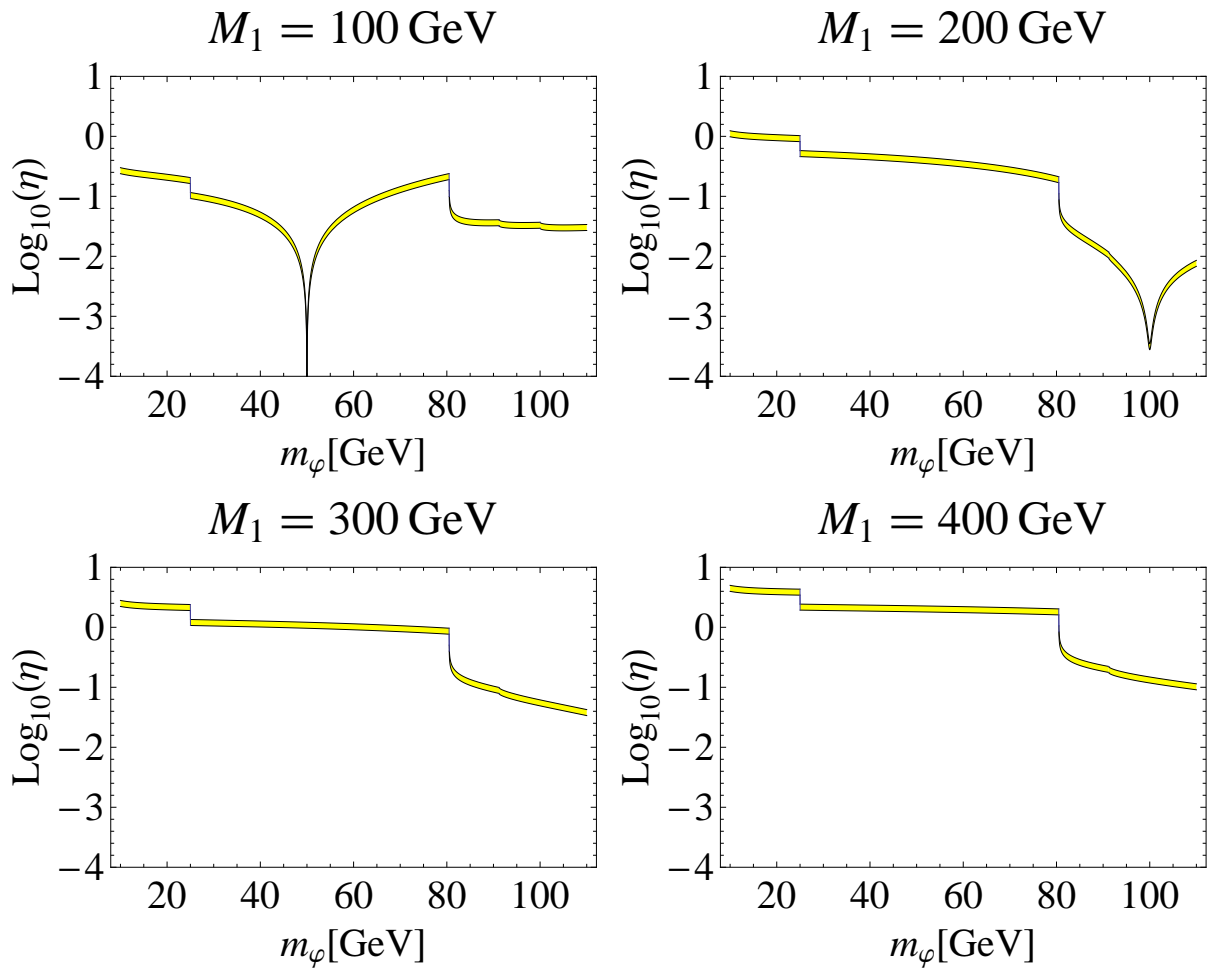}
\caption{Inert-scalar coupling $\eta$ (vs $m_\vp$) required
  by the observed DM abundance $\Omega_{DM}h^2=0.106 \pm
  0.008$~\cite{Amsler:2008zzb} within a 3-$\sigma$ band. As 
  indicated above each panel, the lightest
  Higgs-boson mass ranges from $M_1=100$ to $400\gev$ .}
\label{eta}
\end{figure}

For an estimate of the DM abundance, we will consider two $\vp \vp$
annihilation mechanisms.
First we assume that $\vp \vp$ annihilate to $\gamma\gamma$,
$q\bar{q}$, $l^+l^-$, $W^+W^-$ and $ZZ$ through s-channel $H_1$
exchange.  Then, following \cite{Kolb} we obtain\footnote{This cross
  section is also to be found in the literature~\cite{Burgess:2000yq},
  however our result is smaller by a factor of 2. We have
  included both the combinatoric factor $1/2$ in $\vp\vp H_1$ vertices
  and statistical factors (both in the initial and in the final
  states) in $\langle \sigma v \rangle$.} in the non-relativistic
approximation the following result for the thermally averaged
annihilation cross section
\begin{equation}
\langle \sigma v \rangle_1 = \frac{4 \kappa_1^2 v^2}{(4\mvps-M_1^2)^2+M_1^2
\Gamma_{H_1}^2} \left[\frac{\Gamma_{H_1}(2\mvp)}{2\mvp}\right]
\label{sigv1}
\end{equation}
where $\Gamma_{H_1}(2\mvp)$ stands for the decay width of $H_1$
calculated at $M_1=2 \mvp$ (in the following numerical calculations we
will use the SM width for the estimate).  Now we have to add the
contribution from the $H_1H_1$ final state. There are two
contributions: due to $s$-channel Higgs exchange, and due to the
four-point coupling. We find in the non-relativistic approximation
\begin{equation}
\langle \sigma v \rangle_2 =
\frac{1}{32\pi} \frac{1}{\mvp^2}
\left(1-\frac{M_1^2}{\mvp^2}\right)^{1/2} \theta(\mvp-M_1) \left|
\lambda_{11}
+\frac{\kappa_1 \tilde\lambda_{111}v^2}{4\mvp^2-M_1^2 + i M_1 \Gamma_{H_1}} 
\right|^2,
\label{sigv2}
\end{equation}
where the quartic $\vp^2H_1H_1$ coupling $\lambda_{11}$ is defined by
Eq.~(\ref{Eq:lambda_ij}) and the trilinear $H_1H_1H_1$ coupling
normalized to $v$ is denoted by $\tilde\lambda_{111}$
\cite{Osland:2008aw}.  For an order-of-magnitude estimate of the DM
abundance, we will use here $\tilde\lambda_{111}=3 M_1^2/v^2$ (this
choice, together with (\ref{param}), reproduces results which would be
obtained for the SM Higgs doublet $\phi_{SM}$ coupled to the singlet
through the term $\eta \vp^2 |\phi_{SM}|^2$) and parameterize
$\kappa_1$ and $\lambda_{11}$ through one variable $\eta$ as follows:
\begin{equation}
2 \lambda_{11} = \kappa_1 \equiv \eta.
\label{param}
\end{equation}
Then, following \cite{Kolb} for cold relics one has to solve the
following equation to determine the freeze-out temperature from
$x_f=m_\vp/T_f$:
\begin{equation} 
x_f=\ln\left[0.038\frac{\mpl\mvp}{(g_\star x_f)^{1/2}}
\langle \sigma v \rangle\right],
\end{equation}
where $\langle \sigma v \rangle \equiv \langle \sigma v \rangle_1
+\langle \sigma v \rangle_2 $ and $g_\star$ counts relativistic
degrees of freedom at annihilation and $\mpl$ denotes the Planck
mass. It turns out that in the range of parameters we are interested
in, $x_f\sim \ocal(25)$, so that this is indeed the case of cold dark
matter, it also implies that $g_\star \simeq 10-100$.  Then the
present density of $\vp$'s is given by
\begin{equation}
\Omega_\vp h^2 
= 1.07\cdot 10^9 \frac{x_f}{g_\star^{1/2} \mpl \langle \sigma v \rangle}
\label{om}
\end{equation}
In Fig.~\ref{eta} we show the 3-$\sigma$ allowed band in $\log (\eta)$
vs.\ $m_\vp$, as constrained by the observed DM abundance
$\Omega_{DM}h^2=0.106 \pm 0.008$~\cite{Amsler:2008zzb}.  For
$\mh=100$ and $200\gev$ we observe consequences of resonant behavior
at $\mvp=\mh/2$. The thresholds seen at $\mvp \simeq 25$ and $80\gev$
are caused by the rapid change in $g_{\star}$ as a function of
temperature and by the opening of the $W^+W^-$ channel for the decay
of a Higgs boson of mass $\mh=2m_W$, respectively.

One can conclude that the singlet could indeed provide a realistic
candidate for DM: for any $m_\vp$ between $\sim 1\gev$ and $\sim
500\gev$ there exists an allowed value $\eta$ for which $\Omega_\vp
h^2$ agrees with the experimental data. Note that if we had found only
solutions with $\eta \gsim 1$ and light $\vp$ ($m_\vp \lsim v$) then
this scenario would be jeopardized since the minimization condition
requires $m_\vp^2 > \eta_1 v_1^2 + \eta_2 v_2^2$ (as
$\mu_\vp^2>0$). As seen from Fig.~\ref{eta}, this is not the case.
Most of the allowed region corresponds indeed to $\eta \lsim 10^{-1}$
if the singlet mass is not too low.

\section{Summary}
\label{sum}
The goal of this work was to build a minimal realistic model which
would allow for softening the little hierarchy problem through
suppression of the quadratic divergences in scalar boson mass
corrections and through lifting the mass of the lightest Higgs
boson. That could be accomplished within Two-Higgs-Doublet
Models. Phenomenological consequences of requiring no
quadratic divergences in corrections to scalar masses within the 2HDM
were discussed.  The 2HDM type II was
analyzed taking into account existing experimental
constraints. Allowed regions in the parameter space were
determined. An interesting scalar mass degeneracy was observed for
$\tgb \gsim 40$. The issue of possible CP violation in the scalar
potential was addressed and regions of $\tgb-M_{H^\pm}$ with
substantial strength of CP violation were identified. In order to
accommodate a possibility for dark matter a scalar gauge singlet was
added to the model. Requirements necessary for correct present
abundance of dark matter were estimated.

The model we considered here allows to soften the little hierarchy
problem by lifting the minimal scalar Higgs boson mass and by
suppressing the one-loop quadratic corrections to scalar masses. The
cutoff implied by the naturality arguments is lifted from $\sim
600\gev$ in the SM up to at least $\gsim 2.5 \tev$, depending on the mass of
the lightest scalar.

\vspace{.5in}
\acknowledgments

This work is supported in part by the Ministry of Science and Higher
Education (Poland) as research project N~N202~006334 (2008-11). 
B.G. acknowledges support of the
European Community within the Marie Curie Research \& Training
Networks: ``HEPTOOLS" (MRTN-CT-2006-035505) and ``UniverseNet"
(MRTN-CT-2006-035863).
The research of P.O. has been supported by the Research Council of
Norway.

\end{document}